\documentclass{aa501}
\usepackage{graphics,amssymb,natbib,rotating,psfrag,amsmath,epsfig,array}

\title{Temporal properties of gamma ray bursts as signatures of jets from the central engine}
\titlerunning{Temporal properties of GRBs}

\author{F.\,Quilligan\inst{1} \and
    B.\,McBreen\inst{1} \and
    L.\,Hanlon\inst{1} \and
    S.\,McBreen\inst{1} \and
    K.J.\,Hurley\inst{1} \and
    D.\,Watson\inst{2}
    }

\offprints{fquillig@bermuda.ucd.ie}

\institute{Department of Experimental Physics, University College Dublin, Dublin 4, Ireland
    \and
       X-Ray Astronomy Group, Department of Physics and Astronomy,
              Leicester University, Leicester LE1 7RH, UK}

\date{Received / Accepted}
\abstract{ A comprehensive temporal analysis  has been performed
on the 319 brightest GRBs with $T_{90}$$>$$2\,$s from the BATSE
current catalog. The GRBs were denoised using wavelets and
subjected to an automatic pulse selection algorithm as an
objective way of identifying pulses and quantifying the effects
of neighbouring pulses. The number of statistically significant
pulses selected from the sample was greater than 3000. The rise
times, fall times, full-widths at half maximum (FWHM), pulse
amplitudes and pulse areas were measured and the frequency
distributions are presented here. All are consistent with
lognormal distributions provided the pulses are well separated.
The distribution of time intervals between pulses is not random
but compatible with a lognormal distribution when allowance was
made for the 64 ms time resolution and a small excess (5\%) of
long duration intervals that is often referred to as a
Pareto-L\'{e}vy tail. The time intervals between pulses are most
important because they may be an almost direct measure of the
activity in the central engine. Lognormal distributions of time
intervals also occur in pulsars and SGR sources and therefore
provide indirect evidence that the time intervals between pulses
in GRBs are also generated by rotation powered
systems with super-strong magnetic fields. \\
[\parsep] \indent A range of correlations are presented on pulse
and burst properties. The rise and fall times, FWHM and area of
the pulses are highly correlated with each other.  The pulse
amplitudes are anticorrelated with the FWHM. The time intervals
between pulses and pulse amplitudes of neighbouring pulses are
correlated with each other. It was also found that the number of
pulses, N, in GRBs is strongly correlated with the fluence and
duration and that can explain the well known correlation between
duration and fluence. The GRBs were sorted into three categories
based on N i.e. 3$\leq$N$\leq$12, 13$\leq$N$\leq$24 and
N$\geq$25. The properties of pulses before and after the
strongest pulse were compared for three categories of bursts. No
major differences were found between the distributions of the
pulse properties before and after the strongest pulse in the GRB.
However there is a strong trend for pulses to have slower rise
times and faster fall times in the first half of the burst and
this pattern is strongest for category N. This analysis revealed
that the GRBs with large numbers  of pulses have narrower and
faster pulses and also larger fluences, longer durations and
higher hardness ratios than the GRBs with smaller numbers of
pulses. These results may be explained by either homogeneous or
inhomogeneous jet models of GRBs. The GRBs with larger number of
pulses are closer to the axis if $\Gamma$ varies with the opening
angle of the jet and the imprint of the jet is preserved in the
pulse structure of the burst. The distribution of the number of
pulses per GRB broadly reflects the beaming by the jet.
        \keywords{Gamma rays -- bursts: Gamma rays -- observations: Methods
        -- data analysis: Methods -- statistical}
}

\begin{document}

 \maketitle

\section{Introduction}

Much of the recent progress in the study of gamma-ray bursts
(GRBs) results from the detection of bursts with good location
accuracy by BeppoSAX that enabled the detection of counterparts at
other wavelengths. The subsequent redshift determination of bursts
have established that these bursts are at cosmological distances
\citep{cfpc:1997,vanpara:1997}. GRBs seem to be connected to
massive stars and become powerful probes of the star formation
history of the universe \citep{lamb:2000,han:2000,berger:2001}.
However not many redshifts are known and there is still much work
to be done to determine the mechanisms that produce these
enigmatic events.

The most plausible GRB progenitors are expected to be a newly
formed black hole (BH) surrounded by a temporary accretion disk
\citep{rees:1999,mes:2001,castro:2001}.  The most popular models
include the merger of a neutron star (NS) and a NS
\citep{elp:1989,ruffjan:1999}, NS and a BH \citep{pacz:1991}, BH
white dwarf merger \citep{fryer:1999} and models of failed
supernovae or collapsars \citep{macfad:1999,pacy:1998}.  An
important exception is the model in which the GRB energy is
provided by a newly formed neutron star
\citep{usov:1992,thomp:1994}.  Various explanations have been put
forward for the complicated structure of the light curves. These
range from internal shocks, caused by variations in the velocity
of the outflow \citep{reemes:1994,piran:1999}, to external shocks,
caused by interactions with an external medium
\citep{meszar:1993,derm:1999}. In the internal shock model the
instabilities in the wind leads to shocks which convert a fraction
of the bulk kinetic energy to internal energy remote from the
central engine. A turbulent magnetic field then accelerates
electrons which radiate by synchrotron emission and inverse
Compton scattering, generating the GRB. Many of the observed
features in bursts can be reproduced in the internal shock models
of GRBs
\citep{sapi:1997,kps:1997,daimoc:1998,pansm:1999,downes:2001}.

A variety of analytical techniques has been applied to the
temporal and spectral profiles of GRBs which place  constraints on
the observed distributions which models must satisfy. The
impressive results from these studies include (1) hard to soft
evolution \citep{golens:1983,borgon:2001}; (2) the
duration-hardness anticorrelation \citep{kmf:1993}; (3) the
temporal asymmetry of pulses in GRBs \citep{nnw:1993,lp:1996};
(4) a bimodal duration distribution of GRBs consistent with two
lognormal distributions \citep{kmf:1993,mhlm:1994}; (5) the
discovery of two different types of pulses in
GRBs\citep{ppb:1997}; (6) a correlation between E$_{\rm peak}$
and intensity\citep{mpp:1995}; (7) energy dependence of the pulse
duration \citep{nnb:1996}; (8) a relationship between the pulse
peak energy, E$_{\rm peak}$, and the photon fluence
\citep{lika:1996,crider:1999}; (9) lognormal pulse shapes and
time intervals between pulses in long \citep{mhlm:1994,hmq:1998}
and short GRBs \citep{sheila:2001}; (10) spectra well fit with a
Band function \citep{band:1993}; (11) spectral hardening before a
count rate increase \citep{bhat:1994}; (12) an x-ray excess in
GRB spectra \citep{stroh:1998}; (13) a correlation between
complexity and brightness \citep{stern:1999} and (14) the unique
properties of the pulses and power law relationships between the
pulse properties and durations of GRBs \citep{smcb:2002}

While GRBs display hard to soft spectral evolution, there is
remarkable constancy of the pulses in GRBs throughout the burst
\citep{ramfen:2000,qhm:1999}. The temporal and spectral properties
of a few GRBs with known redshift have yielded two important
results to suggest that GRB properties may be related to their
luminosities. Ramiriz-Ruiz and Fenimore (1999) have shown that
more rapidly variable bursts have higher absolute luminosities.
\citet{nmb:2000} have found an anticorrelation between the time
delay in the arrival times of hard and soft photons in pulses and
the luminosity of the GRB.

The light curves of GRBs are irregular and complex. Statistical
studies are necessary to characterise their properties and hence
to identify the physical properties of the emission mechanism. The
statistical methods used for temporal studies can be broadly
divided into four categories: (1) fits to individual pulses in the
GRB using a number of pulse shape parameters
\citep{nnb:1996,lee:2000,lbp:2000}; (2) a non-parametric approach
to pulse shapes in GRBs
\citep{mhlm:1994,hmq:1998,ymr:1995,qhm:1999}; (3) the average
statistical properties of GRBs using a peak-aligned profile
\citep{ss:1996}; and (4) the average power spectral density of
GRBs \citep{belli:1992,belss:2000,changyi:2000}.  One of the first
studies \citep{mhlm:1994} revealed that lognormal distributions
can adequately describe the properties of GRBs. Subsequent
studies \citep{lifen:1996,hmq:1998,qhm:1999} have confirmed the
applicability of lognormal distributions in accounting for the
wide range in the observed properties of pulses in GRBs.  This
result is not surprising because lognormal distributions arise
from the product of probabilities of a combination of independent
events and such conditions apply to the pulse generation process
in GRBs.

In a different approach \citep{belss:2000} used Fourier analysis
to study the power spectral density of long GRBs. This approach
revealed that the diversity of GRBs is due to random realisations
of the same process which is self-similar over a range of time
scales \citep{ss:1996}.  The slope of the PSD was -5/3 suggesting
that GRBs are related to fully developed turbulence.  The two
different approaches are quite similar because the lognormal
approach has been used to describe fully developed turbulence
\citep{amm:1999}.

The work presented here expands on the earlier analysis
\citep{quilligan:2000} and provides new insight into the mechanism
which generates GRBs.  The aim is to provide a comprehensive
description and understanding of the pulse properties in GRBs and
combine it with other studies of the spectral properties. The
wavelet analysis and the pulse selection algorithm are described
in Sect. 2. The method for comparing the properties of the pulses
before and after the strongest pulse in the GRB is also described
in Sect. 2. The results are presented in Sect. 3, and discussed
in Sect. 4. The conclusions are presented in Sect. 5.

\section{Data Preparation}

The dataset used was taken from the BATSE current catalogue. The
`discsc' files are available at
http://www.batse.msfc.nasa.gov/batse/grb/catalog/4b/
\citep{pmpbk:1999}. The files contain the data from the four
energy channels, which were combined into a single channel to
maximise the signal to noise ratio.  The shapes of GRB pulses
vary little with energy and pulses in different energy channels
can be added together and nearly retain their initial shape.  A
subset of the BATSE catalogue was selected based on the criteria
\citep{nnb:1996} that the GRB duration was greater than two
seconds ($T_{90}>2$\,s) and the peak flux $P_{\rm 256 ms}
> 3.28$ photons cm$^{-2}$s$^{-1}$.  In this way a sample of 324 bursts
with good signal to noise and clearly resolved features was
obtained. Five of these bursts could not be analysed properly due
to data gaps, and so our final sample consisted of 319 GRBs.  All
319 GRBs were used for the timing analyses.  The 250 GRBs that
were summed over only two LAD detectors were used for all
analyses involving pulse amplitude and area.

\subsection{Background subtraction}

The first step in the data preparation involved selecting the
appropriate background for subtraction from the GRB. The start and
end times for each burst were identified. A further margin of 10
seconds was added to both the beginning and end of this chosen
section. Two background sections of duration 30\,s were then
selected, one finishing 20\,s before the start of the section
containing the burst and the other starting 20\,s after the end of
the burst (Fig.~1). These two regions were used to fit a linear
background that was subtracted from the burst section.

\begin{figure}
  \begin{center}
\resizebox{\columnwidth}{!}{\rotatebox{-90}{\includegraphics{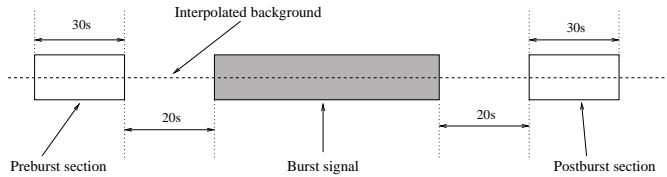}}}
    \caption{Illustration of the background subtraction algorithm. The
shaded region indicates the bursting phase of the GRB.}
  \end{center}
\end{figure}
\subsection{Denoising technique}

One of the difficulties in analysing the time profiles of GRBs is
in overcoming the limitations imposed by the presence of noise in
the signal and the overlap of the individual pulses. The transient
nature of GRBs also means that the usual assumptions for Fourier
transform techniques do not hold \citep{changyi:2000,suzuki:2001}.
An alternative method of filtering the signal is with wavelets.
Wavelet analysis was pioneered by \citet{daubech:1992} and others
during the 1980's \citep{mey:1993}.

Wavelets are specific functions that, when convolved with the signal under
investigation, produce a transformed signal that represents the location and
strength of variations within the original data. The convolution is applied
repeatedly to the data over a range of scales with the convolution function
gradually stretched to coarser and coarser scales, revealing variations at
corresponding scales in the original signal. This iteration with scaling of
the convolved function allows the identification of structure with a variety
of extents both in the spatial and frequency domains.

If the wavelet function is written as $\psi(x)$ then, more formally,
the transform at a particular scale $s$, can be written as
\begin{equation*}
W_{s}f(s,x)=f*\psi_s(x)
\end{equation*}
\noindent where $ \psi_s(x) \equiv (1/s)\psi(x/s)$ represents the
wavelet dilated by a scale factor $s$. At each scale $2^j$, a
discrete wavelet transform that we denote by $W^d_{2^J}$ can be
computed. For the particular choice of scale, $s=2^j, j=1 \ldots
J$ the sequence
\[\left\{ S^d_{2^J},\left( W^d_{2^J} f\right)_{1\leq j\leq J }\right\}\]
is called the discrete dyadic wavelet transform of the input
discrete signal $D=(S_1 f(n))_{n\in Z}$. The $W^d_{2^J} f$
components provide the \textit{detail} at each scale, meaning the
response of the (scaled) wavelet function to the detailed
variation of the signal. The coarse signal, $S^d_{2^J}$, provides
the low frequency (slowly varying) component of the signal
remaining at scales larger than $2^J$. The higher frequency
components can all be recovered from the dyadic wavelet transform
$\left(W^d_{2^j}f\right)_{1\leq j\leq J} $ between scales $2^1$
and $2^J$. A fast algorithm for calculating the wavelet transform
of a signal was developed by \citet{mallzh:1992} and
implementation of this algorithm is at the core of the denoising
procedure used. In Fig.~2, GRB 920513 is shown together with its
dyadic wavelet transform for scales $s=2^{2\ldots4}$ (the lowest
scale, $s=2^1$, is not shown because it is dominated by noise).
The transforms at each of the scales are shown (Fig.~2c, d and
e), along with the low frequency signal $S^d_{2^4}$ containing
the remaining information for scales $i>4$ (Fig.~2f).

The function $\psi(x)$ used in the algorithm was chosen so that
the wavelet acts like an edge detector with a delta response
function to a step edge. In fact $W^d_{2^j}f$ is proportional to
the derivative of the original signal smoothed at the scale $2^j$.
Thus calculating the positions of the modulus maxima of the
transform, $|W^d_{2^J} f|$, is analogous to locating the sharp
variations in the original signal. \citet{mallzh:1992} also
developed an algorithm for allowing the reconstruction of a
signal given just the modulus maxima of the wavelet transforms
across a set of scales along with the low frequency signal
remaining at the coarsest scale. This reduced representation can
reconstruct an accurate copy of the original, using an iterative
algorithm that converges quite quickly to acceptable levels.

A technique for identifying white noise and removing it without
losing any other information was introduced by
\citet{mallhw:1992}. This technique relies on characterising the
behaviour of noise across the various scales in the wavelet
transform using just the information present in the extrema
wavelet representation described in the previous paragraph.  The
change in the amplitude of the extrema between scales allows the
number called the Lipschitz exponent to be calculated. More
precisely, each extremum describes a particular curve in
$\left(\log (s), \log |
  Wf(s,x) |\right)$ space representing its increase or decay on all
the scales for which $Wf(s,x)$ has been computed. Then the
Lipschitz exponent, $\alpha$, is just the maximum slope of a
straight line that remains above this curve \citep{ymr:1995}.
Using results from an analysis of white noise
($\alpha<-\frac{1}{2}$, \citet{mallhw:1992}), as well as the
investigation described below, the characteristic distribution of
Lipschitz exponents for the noise present in BATSE GRB profiles
was determined. In general, noise is expected to have a negative
Lipschitz exponent indicating a decreasing amplitude with
increasing scale. Extrema in the wavelet transform which fall in
the range expected for noise can be removed using an algorithm
developed for this purpose.

As an example consider the extremum at around 110\,s on the top
(finest) scale of GRB 920513 (Fig.~2c). On the next scale, the
amplitude of this extremum is considerably smaller and by the
following scale it is hardly detectable. It is clear that this
extremum corresponds to a noise feature in the original signal and
visual examination of the transform indicates that the amplitude
of the wavelet transform decays quickly when moving to
progressively coarser scales, as expected for noise. The extremum
at around 95\,s corresponds to a pulse with intensity over 25,000
counts per 64 ms. The amplitude of the wavelet maximum increases
on coarser scales, contrary to the behaviour expected for noise.

\begin{figure*}
  \begin{center}

\resizebox{\textwidth}{!}{\includegraphics{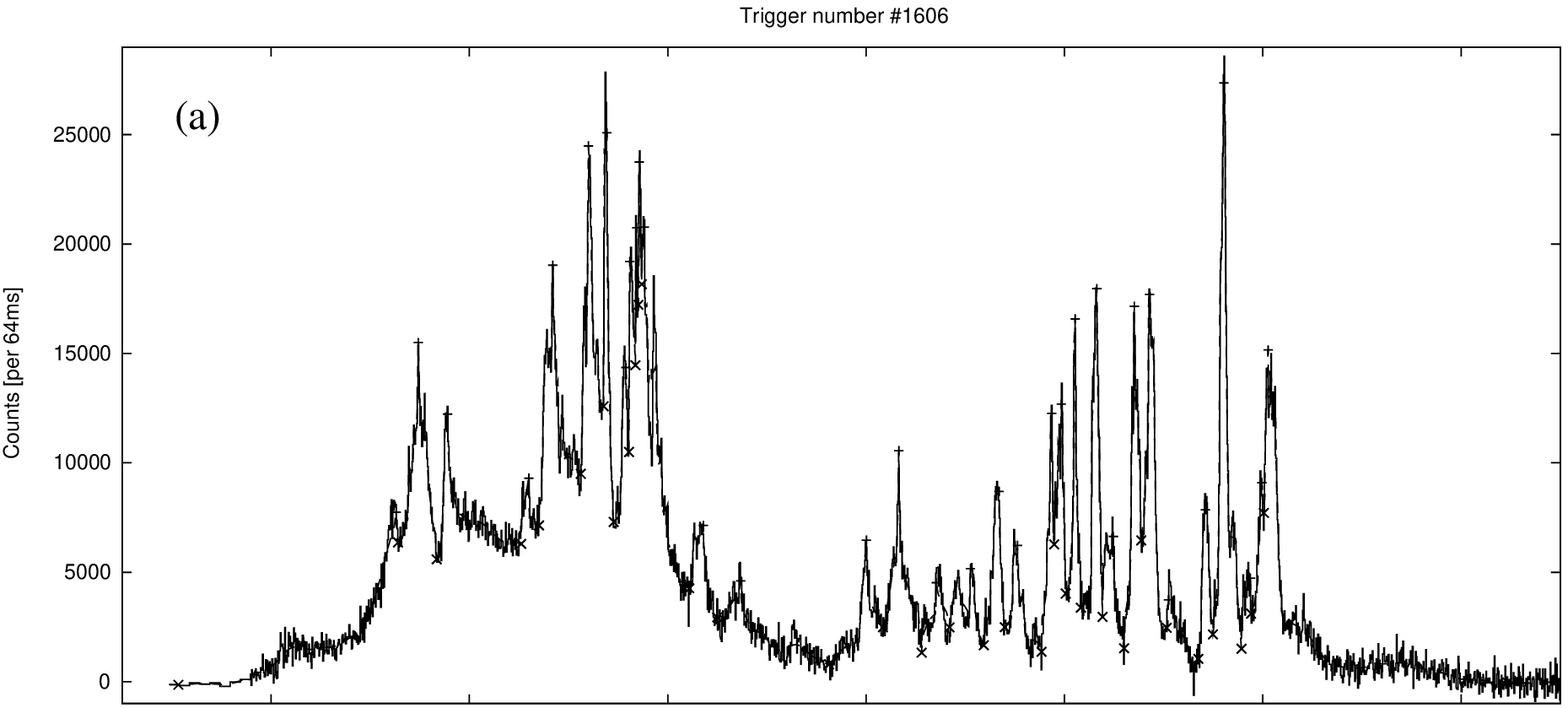}}
\resizebox{.98\textwidth}{!}{\includegraphics{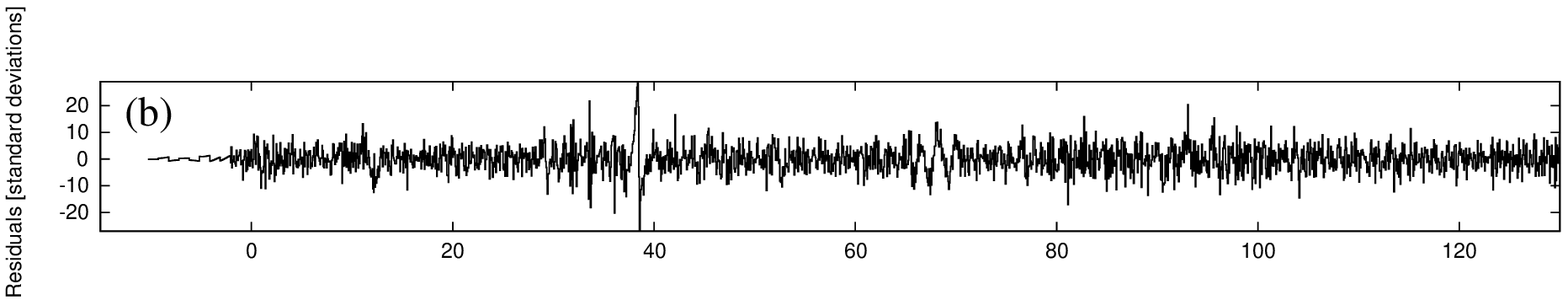}}
\resizebox{0.93\textwidth}{!}{\rotatebox{-90}{\includegraphics{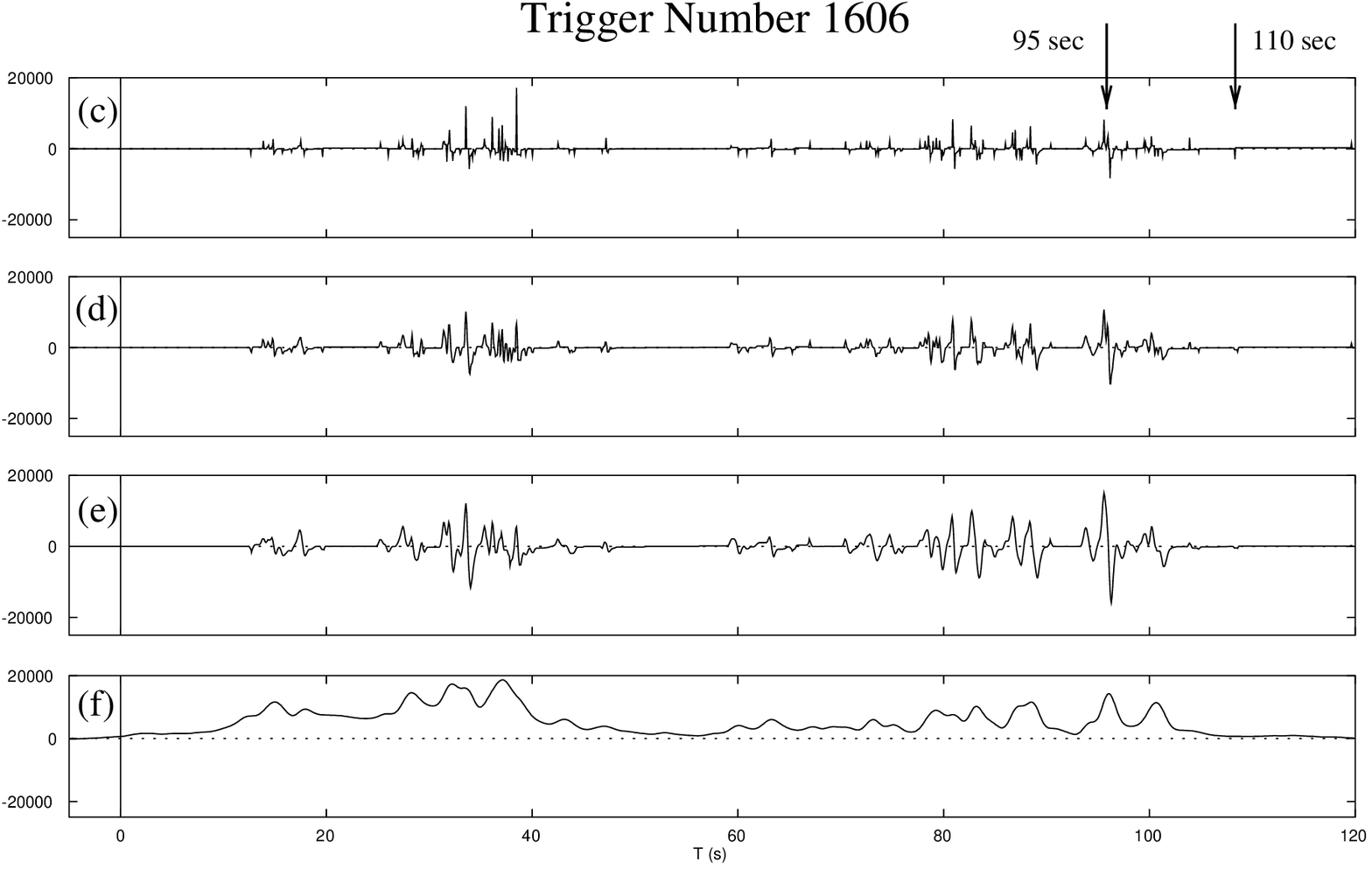}}}
    \caption{The wavelet reconstruction of GRB 920513. (a) The GRB profile
      and wavelet fit with maxima (identified by the + symbol) and minima
      (identified by the x symbol). A total number of 33 pulses were
      identified above a threshold of 5$\sigma$. (b) The residuals between
      the wavelet fit and the actual GRB profile. (c)-(e) The decomposition of GRB 920513
      (BATSE trigger 1606) into its component wavelet scales. The wavelet
      transform for scales $s=2^{2\ldots4}$ (the finest scale, $s=2^1$, is
      not shown because it was dominated by noise). (f) The low
      frequency signal $S^d_{2^4}$ containing the remaining information for
      scales $i>4$.}
   \end{center}
\end{figure*}

The extrema removal algorithm was combined with a simple
thresholding procedure based on the analysis of a section of the
burst background. For each burst being prepared, a section of the
same burst well separated from the signal was used to generate the
wavelet transform of the background. The standard deviation of the
transform at the scale $2^2$ (the finest scale not dominated by
noise) was combined with a significance level ($\sigma_T$) to
calculate an amplitude threshold for the wavelet coefficients of
the signal at the same scale.  Any extrema in the signal with
amplitudes less than this threshold on scale $2^2$ were assumed to
be due to noise and were removed.

A denoised signal was then reconstructed (e.g. Fig.~2a) using the
algorithm described by \citet{mallzh:1992}. The reconstructed
functions have no spurious oscillations or sharp variations, and
are very similar to the original signal. The residuals between
the background subtracted burst and the reconstructed burst are
shown in Fig.~2b. The residuals were calculated assuming that the
errors in the raw data were Poisson distributed. The variation of
the residuals in the section containing signal do not differ
significantly from a section due to background noise (Fig.~2b).

\subsection{Pulse selection}

Each pulse was examined to find minima on each side which were
separated in amplitude from the maximum by more than a chosen
significance level. If the search for minima failed on a pulse it
was rejected and the search for pulses continued. The algorithm
was designed so that the maximum and minima finally selected for a
given pulse were the extreme values in that region of the signal.
The significance level of the maxima and minima was calculated by
multiplying the error on the counts by a scale factor,
$\tau_\sigma$. Given two data points $n_1$ and $n_2$, where
$n_1>n_2$, the points overlap if

\[(n_1-\tau_{\sigma} \Delta n_1)<(n_2+\tau_{\sigma} \Delta n_2)\]

The errors on the counts in the cleaned signal were assumed to be
Poissonian.

The pulse selection process is illustrated in Fig.~3. Starting
with pulse 2, minima 1 and 3 were easily identified. Next, pulse 4
was considered. Pulse 4 overlaps the adjacent minimum 3 (based on
appropriately sized error bars) and both turning points were
rejected and minimum 5 was considered as the true minimum
associated with pulse 2. A minimum, maximum, minimum triplet
consisting of points 1, 2 and 5 was found.  At the conclusion of
the analysis the overlapping maxima/minima pairs consisting of
points 6 and 7, 9 and 10 and 13 and 14 were rejected. The
algorithm selected points 2, 8 and 12 as the maxima and identified
associated minima 1, 5, 11 and 15.

The analysis was then extended to allow the identification of
pulses that were well separated from their neighbours such that
the overlap from the surrounding signal did not significantly
affect the profile of the pulse. The fraction of the total height,
$H_b$, (from pulse to background) which was above the higher
minimum was determined and a threshold was applied to this
fraction, above which pulses were considered isolated. In Fig.~3
the pulses 8 and 12 are not very well separated from their
neighbours whereas pulse 2 is effectively isolated and not
strongly influenced by surrounding signal. The fractional isolated
height of pulse 8 (to which the threshold is applied) was obtained
using (Fig.~3):

\( f_I=\min (H_l,H_r)/H_b = H_r/H_b = 8/21 \approx 0.38, \)

\noindent where $H_l$ is the height difference between the pulse 8
and the point 5 on the left and $H_r$ is the height difference
between the pulse 8 and the point 11 on the right.

For pulse 2 (with point 1 being the higher minimum the smaller
height difference is on the left) the estimate is:

\(
f_I=\min(H_l,H_r)= H_l/H_b = 15/17 \approx 0.88.
\)

Thus, if the threshold, $\tau_I$, were set above $\sim 40$\% then
pulse 2 would be accepted as isolated and pulse 8 would be
rejected. This method provided an objective way to identify and
quantify pulses that were influenced by neighbouring signals.

\begin{figure}
  \begin{center}

\resizebox{\columnwidth}{!}{\rotatebox{-90}{\includegraphics{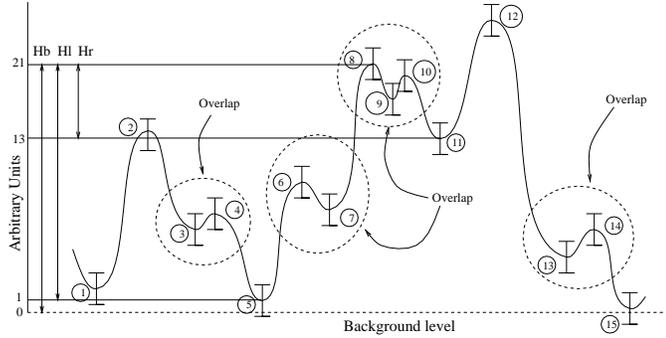}}}
    \caption{A diagramatic representation  of the algorithm for identifying
      pulses with appropriate significance. The error bars were calculated
      using the count rate in the particular bin, multiplied by the
      threshold significance level, $\tau_{\sigma}$. The dashed circles
      highlight maxima and minima where the error bars overlap. Larger
      values of $\tau_{\sigma}$ eliminate overlapping regions that were more
      widely separated.  With a slightly smaller choice of $\tau_{\sigma}$ in
      this example, pulses 6 and 14 would have been accepted.}

  \end{center}
\end{figure}

\subsection{Properties of the pulses}

The next task was to characterise the GRB profiles based on the
properties of the constituent pulses. The following
characteristics which had been studied previously
\citep{mhlm:1994,lifen:1996,hmq:1998} were investigated: the
number of pulses per burst,\, $N$; the time intervals between
pulses,\, $\Delta$T; the pulse amplitudes,\, $C$; the pulse
area,\, $A_p$; the rise and fall times,\, $t_r$ and \,$t_f$; and
the pulse durations or full width at half maximum, FWHM;

The total number of pulses in the sample of 319 GRBs was
determined for a range of thresholds $\tau_\sigma$ and varying
isolation levels, $\tau_I$ (Fig.~4). The variation in isolation
level has a much larger effect on the sample than the threshold,
and caused a reduction in the number of pulses from over 3000 at
the isolation level of 20\% to under 800 at the 80\% level. Fig.~4
also shows that the number of pulses falls quite quickly as
$\tau_\sigma$ increases from 3 to 5. The initial rapid reduction
in the number of pulses selected may be an indication of the
removal of the small population of noise pulses remaining after
the denoising process.  The total number of pulses is not very
sensitive to the threshold level in the region of 5$\sigma$. The
319 GRBs are listed in Table~1, along with the total number of
pulses above 5$\sigma$ for each burst. Also included in the table
are the number of isolated pulses at and above the 50\% and 75\%
levels.

\begin{figure}
    \leavevmode
    \psfrag{xlabel}[b]{\small $\tau_I$}
    \psfrag{ylabel}[t]{\small $\tau_\sigma$}
    \psfrag{zlabel}[t]{\small No. of Peaks}
\includegraphics[width=0.95\columnwidth]{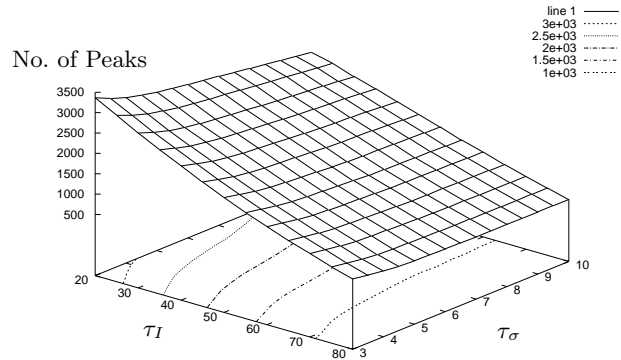}
    \caption{The number of pulses selected from the 319 GRBs as a function
      of the isolation level $\tau_{I}$ and the threshold $\tau_{\sigma}$. The
      plot also shows a projection of the data on to the
      $\tau_{I},\tau_{\sigma}$ plane and the contour levels are as given in the
      legend.}
\end{figure}

\begin{figure}
\resizebox{\columnwidth}{!}{\rotatebox{90}{\includegraphics[width=0.8\columnwidth]{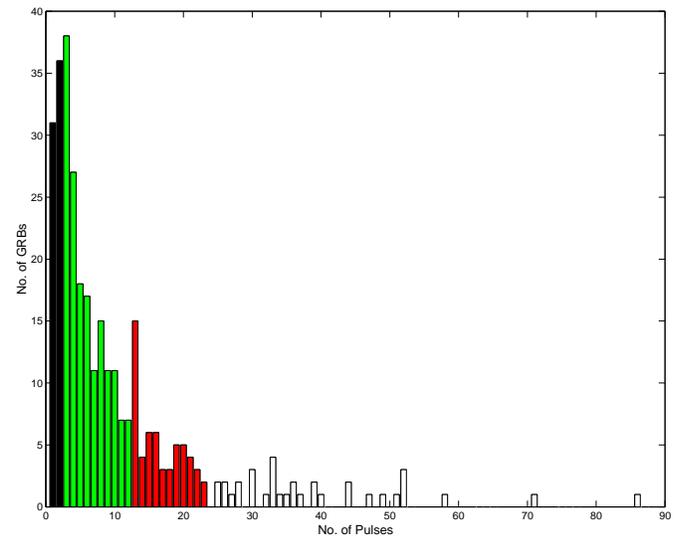}}}
    \caption{The distribution of the number of pulses (N) per GRB.
     The shaded regions highlight the division into four
     categories, namely M, N, O and P with 1 $\leq N \leq 2$, 3 $\leq N \leq 12$,
     13 $\leq N \leq 24$ and N $\geq$ 25 respectively.}
\end{figure}

In the analysis of pulse shapes, non-parametric methods were used
to estimate the various characteristics of the pulse profiles.
This approach was chosen to make the conclusions more robust since
no assumptions were made about the pulse shapes. Also since the
measurements are made on the isolated pulses selected by the
algorithm, the degree of isolation can be varied arbitrarily. If a
particular measurement was sensitive to influence from surrounding
pulses then the threshold $\tau_I$ was increased until the
influence was reduced sufficiently, with the proviso that the
number of pulses in the sample remains statistically useful.

The classification of pulses into isolated and non-isolated
categories based on the algorithm allowed the measurement of
characteristics of the temporal profile which are affected by
neighbouring signals. The level of interference between pulses and
the surrounding signal is dependent on the threshold at which the
selection of these pulses is made. In fact a broad range of
threshold levels were used to examine the time profiles and
$\tau_I$ was typically varied from 20\% to 80\%.  It was decided
based on these results to adopt pulses with $\tau_{\sigma} \geq 5
\sigma$ and $\tau_{I} \geq 50\%$ for the main analysis of the
pulse properties.

The pulse amplitude was measured as the maximum count in a 64 ms
time interval after background subtraction. The pulse area was
measured using the sum of the background subtracted count rates
starting at 5\% of the height of the pulse above the left minimum
to 5\% of the pulse height above the right minimum on the falling
edge of the profile. The starting point at 5\% above the minimum
was chosen to eliminate contributions from background noise for
pulses with minima widely separated from the maximum.

For similar reasons the rise time was measured from 5\% of the
height of the pulse above the left minimum to 95\% of that height.
The upper value of 95\% ensures that the finishing point is robust
against flat topped pulses and noise in the profile near the
maximum. The fall time was measured in a similar way to the rise
times i.e. from 95\% to 5\% of the pulse height above the right
minimum.

The duration of the individual pulses was measured using the FWHM
of the pulse. This approximation is valid only for well isolated
pulses and tended to give poorer estimates for the true pulse
width as the effect of neighbouring signals increased and the left
and right minima of the pulse rose out of the background.

\begin{table*}
 \caption{The BATSE trigger number of the GRBs used in the analysis. N is the total
number of pulses in the GRB, N(50/75) are the numbers of pulses
that are isolated at and above the 50\% and 75\% levels
respectively.}
 \setlength{\tabcolsep}{2pt}
 \setlength{\extrarowheight}{-3pt}
\scriptsize{
\begin{tabular}{@{}rcl rcl rcl rcl rcl rcl@{}}
\hline\hline\\[-3pt]
GRB    &N    &N(50/75) &GRB    &N    &N(50/75) &GRB    &N    &N(50/75) &GRB    &N    &N(50/75) &GRB    &N    &N(50/75) &GRB &N    &N(50/75)\\
\hline\\[-3pt]
105 & 4 & 3/2   & 1974  & 2 & 0/0   & 2994  & 36    & 20/6  & 3929  & 3 & 2/1   & 6113  & 5 & 3/1   & 7318  & 12    & 5/3\\
109 & 19    & 10/2  & 1997  & 13    & 8/2   & 3001  & 3 & 3/2   & 3930  & 19    & 6/1   & 6124  & 30    & 14/5  & 7329  & 3 & 0/0\\
130 & 11    & 5/1   & 2037  & 6 & 5/3   & 3035  & 21    & 4/1   & 3936  & 5 & 1/1   & 6168  & 2 & 1/1   & 7343  & 11    & 2/1\\
143 & 13    & 6/3   & 2053  & 1 & 1/1   & 3039  & 5 & 3/1   & 3954  & 1 & 1/1   & 6198  & 10    & 1/1   & 7360  & 5 & 4/1\\[5pt]
179 & 4 & 3/2   & 2067  & 8 & 1/1   & 3042  & 19    & 12/6  & 4039  & 33    & 21/7  & 6235  & 6 & 3/1   & 7374  & 1 & 1/1\\
219 & 13    & 4/3   & 2080  & 26    & 12/2  & 3057  & 52    & 3/1   & 4048  & 9 & 3/3   & 6242  & 3 & 3/3   & 7429  & 3 & 1/1\\
222 & 6 & 0/0   & 2083  & 2 & 2/1   & 3067  & 8 & 4/3   & 4312  & 4 & 2/1   & 6251  & 3 & 3/2   & 7446  & 1 & 1/1\\
249 & 21    & 4/1   & 2090  & 15    & 7/2   & 3105  & 30    & 19/9  & 4368  & 12    & 2/2   & 6266  & 9 & 2/1   & 7464  & 3 & 2/1\\[5pt]
394 & 25    & 13/3  & 2110  & 13    & 3/1   & 3110  & 13    & 8/3   & 4556  & 3 & 1/1   & 6274  & 18    & 16/6  & 7475  & 2 & 2/2\\
451 & 2 & 2/2   & 2138  & 6 & 2/2   & 3115  & 9 & 6/3   & 4701  & 20    & 8/4   & 6303  & 1 & 1/1   & 7477  & 9 & 3/2\\
467 & 3 & 1/1   & 2151  & 3 & 3/3   & 3128  & 36    & 13/1  & 4814  & 2 & 1/1   & 6321  & 3 & 3/3   & 7491  & 40    & 11/1\\
469 & 1 & 1/1   & 2156  & 39    & 8/3   & 3138  & 4 & 2/1   & 5080  & 7 & 7/3   & 6329  & 9 & 1/1   & 7503  & 4 & 1/1\\[5pt]
503 & 1 & 1/1   & 2213  & 5 & 5/2   & 3178  & 6 & 3/1   & 5304  & 14    & 0/0   & 6335  & 2 & 2/2   & 7527  & 2 & 2/2\\
543 & 3 & 2/1   & 2228  & 28    & 12/5  & 3227  & 17    & 5/3   & 5389  & 13    & 7/3   & 6336  & 3 & 0/0   & 7529  & 1 & 1/1\\
647 & 3 & 1/1   & 2232  & 14    & 8/6   & 3241  & 15    & 10/3  & 5417  & 2 & 0/0   & 6397  & 1 & 1/1   & 7530  & 2 & 2/2\\
660 & 2 & 2/1   & 2316  & 1 & 1/1   & 3245  & 71    & 24/6  & 5419  & 2 & 1/1   & 6400  & 3 & 3/3   & 7549  & 33    & 6/2\\[5pt]
676 & 4 & 2/1   & 2321  & 11    & 4/1   & 3255  & 12    & 6/3   & 5439  & 4 & 4/3   & 6404  & 13    & 7/2   & 7560  & 12    & 8/6\\
678 & 52    & 21/6  & 2329  & 20    & 3/1   & 3269  & 15    & 13/7  & 5447  & 2 & 1/1   & 6413  & 8 & 7/3   & 7575  & 30    & 10/4\\
829 & 3 & 0/0   & 2362  & 2 & 2/2   & 3287  & 10    & 6/1   & 5450  & 13    & 5/4   & 6422  & 2 & 0/0   & 7605  & 9 & 9/3\\
841 & 6 & 6/3   & 2367  & 3 & 3/2   & 3290  & 6 & 4/3   & 5451  & 2 & 0/0   & 6436  & 3 & 3/3   & 7607  & 8 & 8/3\\[5pt]
869 & 10    & 4/1   & 2371  & 2 & 2/1   & 3306  & 16    & 13/4  & 5470  & 7 & 4/4   & 6451  & 3 & 2/1   & 7678  & 37    & 5/2\\
907 & 4 & 3/2   & 2387  & 1 & 1/1   & 3330  & 13    & 5/3   & 5473  & 39    & 27/19 & 6453  & 25    & 5/1   & 7688  & 14    & 9/3\\
973 & 3 & 1/1   & 2393  & 2 & 1/1   & 3345  & 4 & 1/1   & 5477  & 12    & 9/5   & 6472  & 47    & 18/5  & 7695  & 27    & 16/5\\
999 & 2 & 2/2   & 2431  & 1 & 1/1   & 3351  & 10    & 7/5   & 5486  & 12    & 1/1   & 6525  & 8 & 6/5   & 7711  & 1 & 1/1\\[5pt]
1025    & 4 & 2/1   & 2436  & 11    & 8/2   & 3408  & 44    & 17/5  & 5489  & 16    & 8/4   & 6528  & 8 & 3/1   & 7766  & 22    & 15/6\\
1085    & 2 & 0/0   & 2446  & 4 & 3/2   & 3415  & 14    & 8/4   & 5512  & 5 & 1/1   & 6560  & 20    & 16/9  & 7770  & 3 & 2/1\\
1122    & 10    & 3/1   & 2450  & 22    & 7/4   & 3436  & 5 & 4/1   & 5523  & 2 & 2/2   & 6576  & 18    & 11/5  & 7775  & 1 & 1/1\\
1141    & 9 & 0/0   & 2533  & 49    & 9/1   & 3458  & 8 & 2/2   & 5526  & 34    & 34/14 & 6587  & 28    & 9/4   & 7781  & 4 & 2/2\\[5pt]
1157    & 10    & 8/2   & 2537  & 5 & 2/1   & 3480  & 1 & 1/1   & 5530  & 7 & 5/1   & 6593  & 20    & 13/2  & 7788  & 21    & 14/6\\
1159    & 1 & 1/1   & 2586  & 11    & 7/7   & 3481  & 11    & 1/1   & 5548  & 9 & 8/4   & 6621  & 2 & 2/2   & 7845  & 17    & 15/11\\
1190    & 4 & 2/2   & 2611  & 3 & 2/2   & 3488  & 6 & 6/5   & 5563  & 1 & 1/1   & 6629  & 6 & 4/3   & 7858  & 5 & 0/0\\
1204    & 3 & 3/3   & 2628  & 7 & 5/2   & 3489  & 12    & 2/2   & 5567  & 5 & 1/1   & 6630  & 2 & 1/1   & 7884  & 21    & 15/6\\[5pt]
1385    & 23    & 4/1   & 2700  & 4 & 3/1   & 3491  & 3 & 1/1   & 5568  & 3 & 0/0   & 6665  & 8 & 2/1   & 7906  & 13    & 7/5\\
1419    & 3 & 1/1   & 2736  & 1 & 1/1   & 3512  & 1 & 1/1   & 5572  & 3 & 3/3   & 6672  & 4 & 3/1   & 7929  & 6 & 4/1\\
1425    & 7 & 5/1   & 2790  & 20    & 10/1  & 3516  & 10    & 7/2   & 5575  & 6 & 2/1   & 6683  & 10    & 10/3  & 7954  & 16    & 14/13\\
1440    & 15    & 9/2   & 2793  & 7 & 4/3   & 3523  & 16    & 0/0   & 5591  & 10    & 7/4   & 6694  & 10    & 6/1   & 7969  & 2 & 1/1\\[5pt]
1443    & 2 & 2/2   & 2797  & 3 & 1/1   & 3569  & 4 & 2/2   & 5593  & 1 & 1/1   & 6764  & 8 & 3/1   & 7987  & 6 & 2/1\\
1468    & 19    & 17/6  & 2798  & 11    & 1/1   & 3593  & 13    & 6/1   & 5601  & 1 & 1/1   & 6814  & 1 & 1/1   & 7994  & 13    & 1/1\\
1533    & 32    & 20/12 & 2799  & 8 & 6/6   & 3598  & 3 & 3/2   & 5614  & 2 & 2/2   & 6816  & 8 & 7/3   & 7998  & 4 & 3/2\\
1541    & 33    & 11/1  & 2812  & 15    & 12/9  & 3634  & 7 & 7/3   & 5621  & 6 & 3/1   & 6824  & 2 & 2/2   & 8008  & 6 & 3/3\\[5pt]
1578    & 3 & 0/0   & 2831  & 58    & 12/2  & 3648  & 3 & 3/1   & 5628  & 9 & 4/1   & 6904  & 1 & 1/1   & 8019  & 2 & 2/1\\
1606    & 33    & 15/4  & 2852  & 23    & 5/1   & 3649  & 2 & 2/2   & 5644  & 4 & 1/1   & 6930  & 2 & 2/2   & 8022  & 5 & 5/4\\
1609    & 8 & 2/1   & 2855  & 13    & 2/1   & 3658  & 4 & 2/1   & 5654  & 5 & 3/2   & 6963  & 17    & 10/2  & 8030  & 7 & 6/3\\
1625    & 15    & 2/1   & 2856  & 86    & 19/3  & 3663  & 9 & 7/3   & 5704  & 9 & 8/4   & 7028  & 5 & 3/2   & 8050  & 1 & 1/1\\[5pt]
1652    & 6 & 1/1   & 2889  & 26    & 12/4  & 3765  & 4 & 2/2   & 5711  & 2 & 2/1   & 7113  & 52    & 8/1   & 8063  & 4 & 4/3\\
1663    & 16    & 0/0   & 2891  & 18    & 15/4  & 3776  & 7 & 1/1   & 5725  & 13    & 12/6  & 7170  & 16    & 6/2   & 8087  & 8 & 7/3\\
1664    & 5 & 1/1   & 2894  & 5 & 1/1   & 3788  & 4 & 3/1   & 5726  & 3 & 3/1   & 7172  & 1 & 1/1   & 8098  & 4 & 2/2\\
1676    & 35    & 21/7  & 2913  & 5 & 2/2   & 3860  & 13    & 9/3   & 5773  & 4 & 1/1   & 7185  & 19    & 16/7  & 8099  & 1 & 1/1\\[5pt]
1683    & 6 & 3/1   & 2919  & 3 & 3/2   & 3866  & 1 & 1/1   & 5867  & 7 & 6/3   & 7240  & 4 & 4/4   & 8111  & 2 & 2/1\\
1709    & 2 & 0/0   & 2929  & 44    & 16/5  & 3870  & 1 & 1/1   & 5955  & 4 & 2/2   & 7247  & 2 & 1/1   &   &   & \\
1711    & 8 & 2/1   & 2953  & 6 & 2/2   & 3891  & 3 & 2/1   & 5989  & 8 & 7/5   & 7255  & 5 & 1/1   &   &   & \\
1815    & 10    & 5/2   & 2958  & 2 & 2/2   & 3893  & 3 & 1/1   & 5995  & 7 & 1/1   & 7290  & 1 & 1/1   &   &   & \\[5pt]
1883    & 1 & 1/1   & 2984  & 22    & 16/6  & 3905  & 4 & 4/1   & 6090  & 3 & 1/1   & 7301  & 51    & 13/7  &   &   & \\
1886    & 5 & 0/0   & 2988  & 2 & 2/2   & 3912  & 1 & 1/1   & 6100  & 3 & 0/0   & 7305  & 3 & 3/3   &   &   & \\[3pt]
\hline
\end{tabular}
}
\end{table*}

\normalsize{
\subsection{Number of pulses per GRB}

The frequency distribution of the number of pulses (N) per GRB is
given in Fig.~5.  N has a range from 1 to 86 with a peak at a
value of 3, a median of 6 and only 10\% of GRBs have N $\geq$ 25.
For convenience N is divided into four categories (Fig.~5).  There
are 34 GRBs in category P and only 7 have N $>$ 50.  Many of the
timing studies on GRBs have concentrated on the categories with
large N (e.g.\ Norris et al. 1996; Li \& Fenimore 1996)\nocite{nnb:1996,lifen:1996}, which means that these analyses have
focussed solely on the tail of the distribution shown in Fig.~5.

\section{Results}

\subsection{The Lognormal distribution}

It has been shown previously that the lognormal distribution can
adequately describe the properties of GRBs \citep{qhm:1999}. The
lognormal distribution is generated by statistical processes
whose results depend on a product of probabilities arising from a
combination of events \citep{aitbr:1957}. A positive random
variable X is said to be lognormally distributed if Y=log(X) is
normally distributed with mean $\mu$ and variance $\sigma^2$. The
probability density function is:

\begin{equation}
f(x)= \left \{ \begin{array}{ll} {\frac{1}{\sqrt{2\pi} \sigma
x}}\exp{(-(\ln{x}-\mu)^2/2\sigma^2)} & x > 0 \\ 0 & x \leq 0 \\
\end{array}
\right.
\end{equation}

The median of the distribution occurs at $x=e^\mu$. Many examples
of lognormal distributions occur in nature, such as the
propagation of a laser beam in a turbulent medium, the size of
cumulus clouds in the atmosphere and terrestrial lightning
\citep{uman:1987,mhlm:1994}. In the case of terrestrial lightning,
the durations, peak currents, intervals between the strokes in the
flashes, and the flash charges are all lognormally distributed.
The statistical properties of strokes in lightning discharges in
the Earths atmosphere have a close resemblance to the statistical
properties of pulses in GRBs.  In both cases the result depends on
a multiplicative process arising from a combination of independent
events.

\begin{figure}
\includegraphics{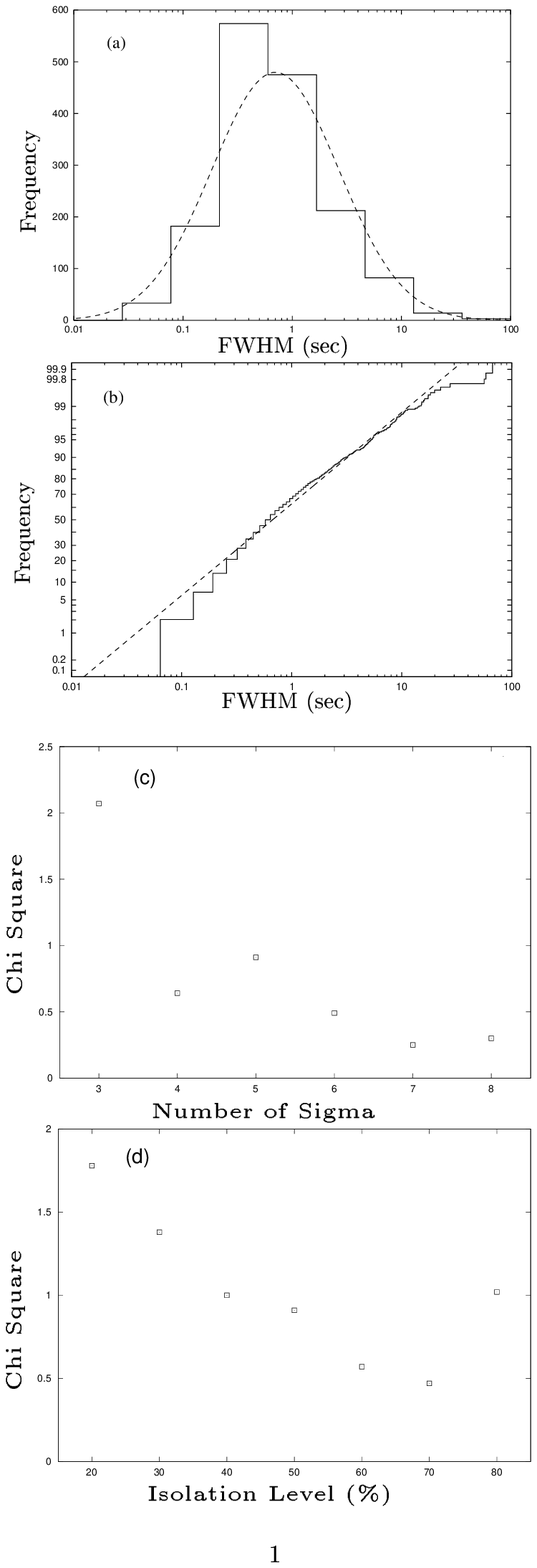}

    \caption{(a) The distribution of the FWHM of pulses with $\tau_{\sigma} \geq 5$ and $\tau_{I} \geq 50\%$ and the best lognormal fit to the data.
    (b) The same
    data plotted as a cumulative percent such that a
    lognormal distribution yields a straight line.  The large
    count in the first bin is due to the 64 ms time resolution.
    (c-d) The values of reduced $\chi^{2}$ for the best lognormal fit as a
    function of $\tau_{\sigma}$ and $\tau_{I}$.}
\end{figure}

\begin{figure}
    \psfrag{xlab}[b]{\small $t_{\rm r}$ (sec)}
        \psfrag{ylab}[b]{\small Frequency}
    \vspace{1em}
\includegraphics[width=0.85\columnwidth]{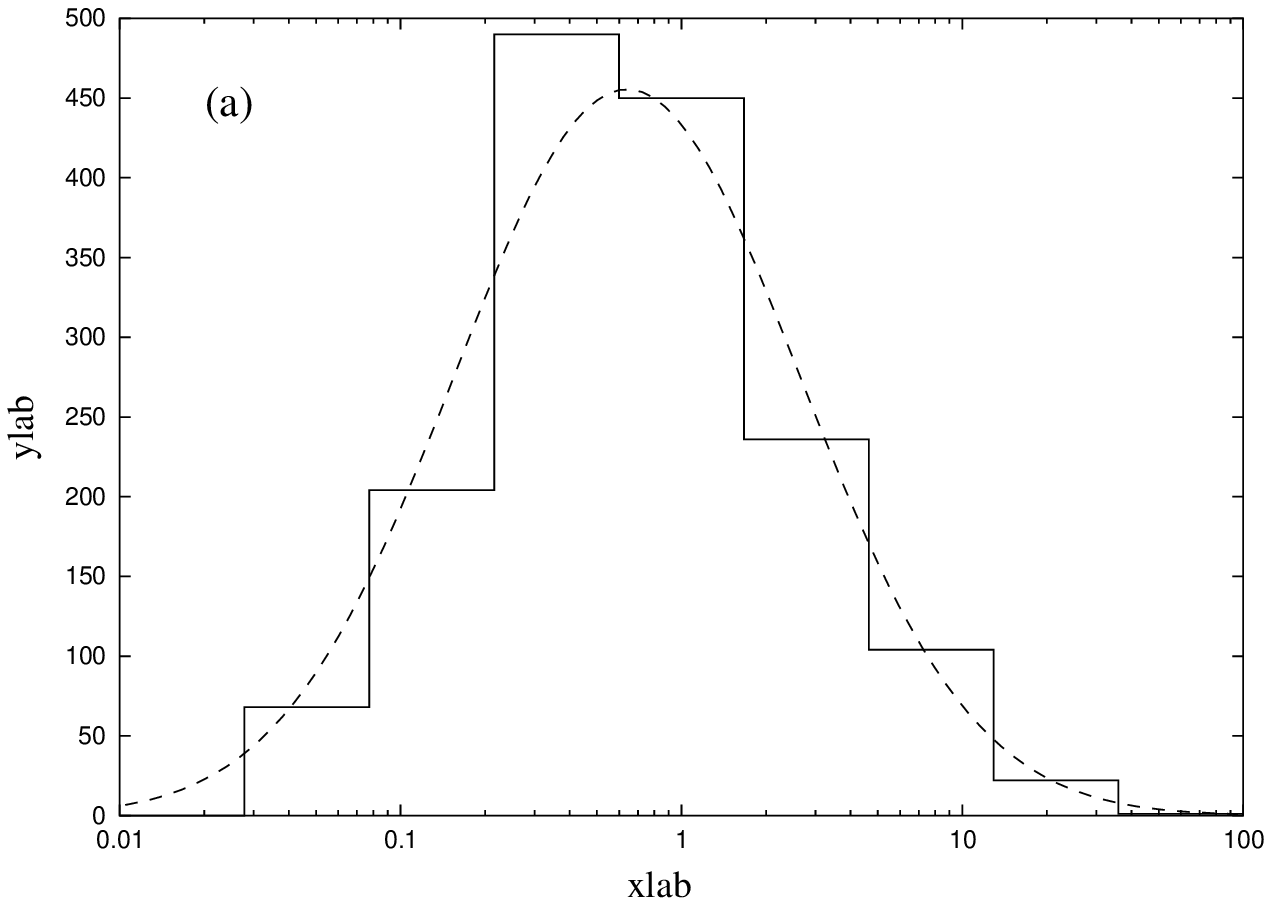}
    \psfrag{xlab}[b]{\small $t_{\rm r}$ (sec)}
        \psfrag{ylab}[b]{\small Frequency}
    \vspace{1em}
\includegraphics[width=0.85\columnwidth]{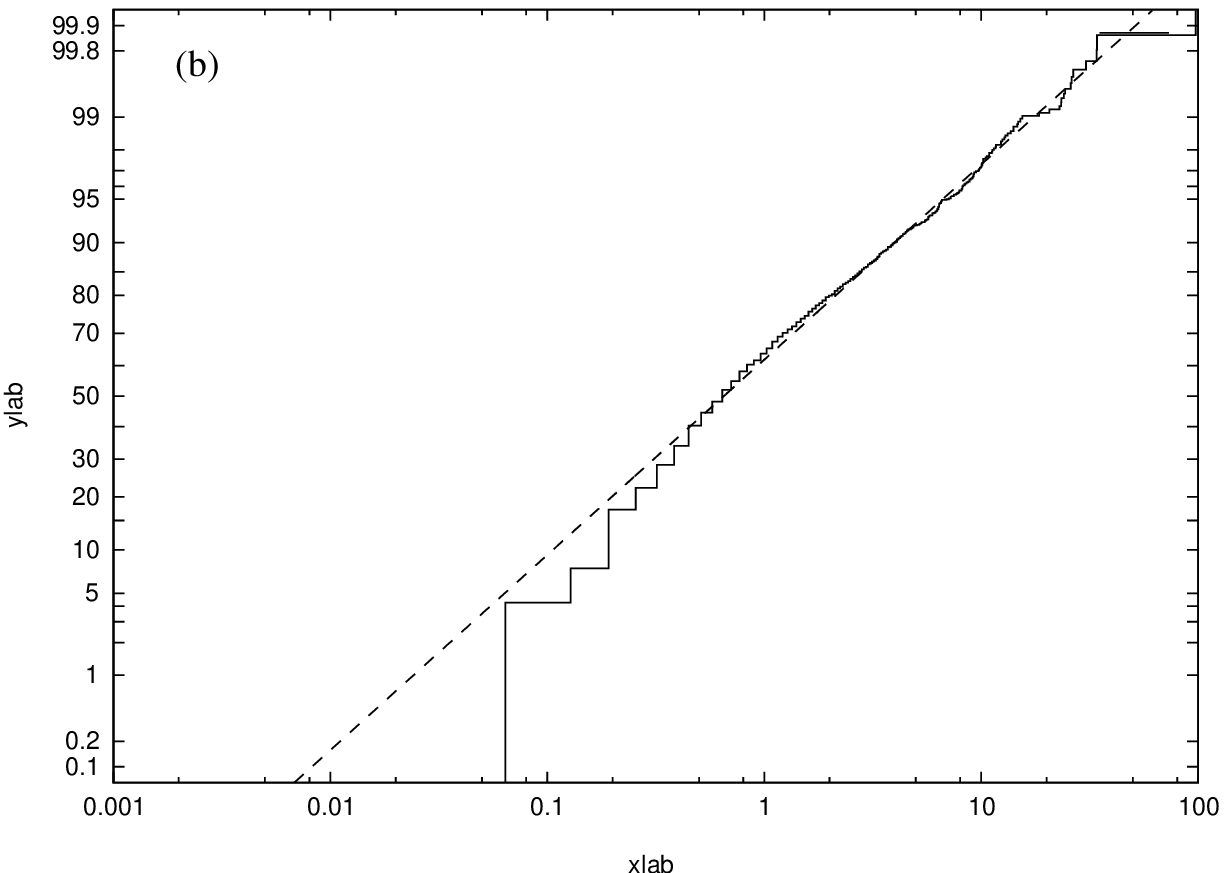}
    \caption{%
    (a) The distribution of the pulse rise times with
    $\tau_{\sigma} \geq 5$ and $\tau_{I} \geq 50\%$ and the best lognormal fit to the data. (b) The same
    data plotted as a cumulative percent.  The large count
    in the first bin is due to the 64 ms time resolution.}
\end{figure}

\begin{figure}
        \psfrag{xlab}[b]{\small $t_{\rm f}$ (sec)}
        \psfrag{ylab}[b]{\small Frequency}
    \vspace{1em}
\includegraphics[width=0.85\columnwidth]{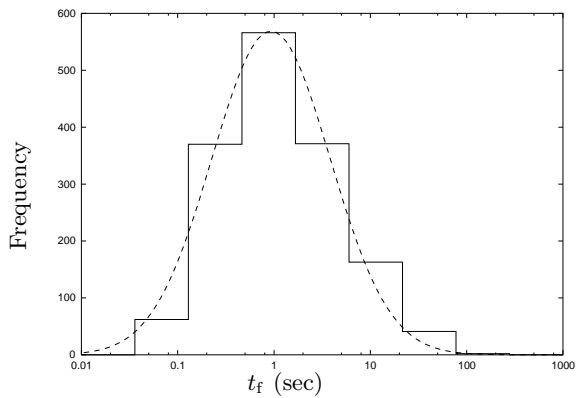}

    \caption{%
    The distribution of the fall times of the pulses with $\tau_{\sigma} \geq 5$ and
    $\tau_{I} \geq 50\%$ and the best lognormal fit to the data.}
\end{figure}

\begin{figure}
        \psfrag{xlab}[b]{\small Pulse Amplitudes (counts)}
        \psfrag{ylab}[b]{\small Frequency}
    \vspace{1em}
\includegraphics[width=0.85\columnwidth]{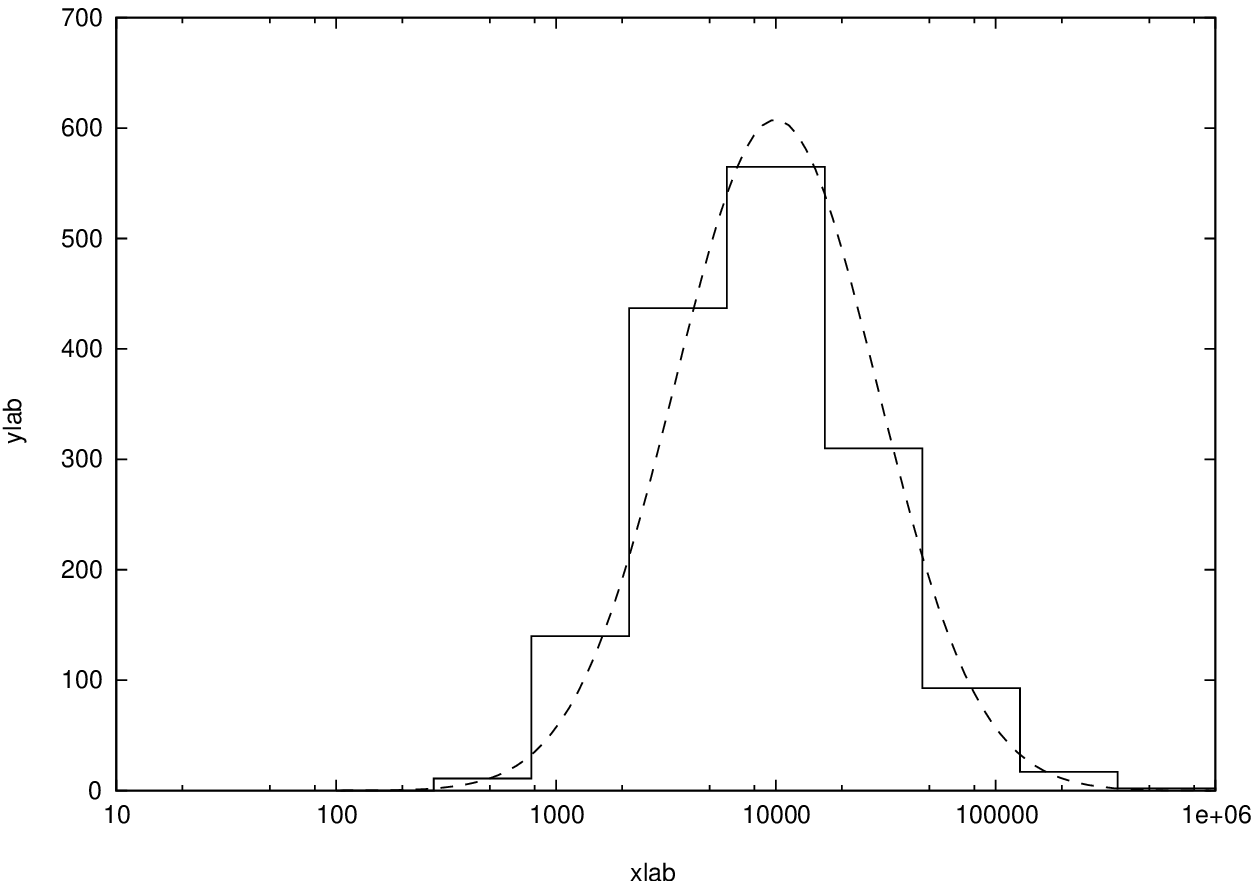}
    \caption{%
    The distribution of the pulse amplitudes with $\tau_{\sigma} \geq 5$ and
    $\tau_{I} \geq 50\%$ and the best lognormal fit to the data.}
\end{figure}

\begin{figure}
        \psfrag{xlab}[b]{\small Pulse Areas (counts)}
        \psfrag{ylab}[b]{\small Frequency}
    \vspace{1em}
\includegraphics[width=0.85\columnwidth]{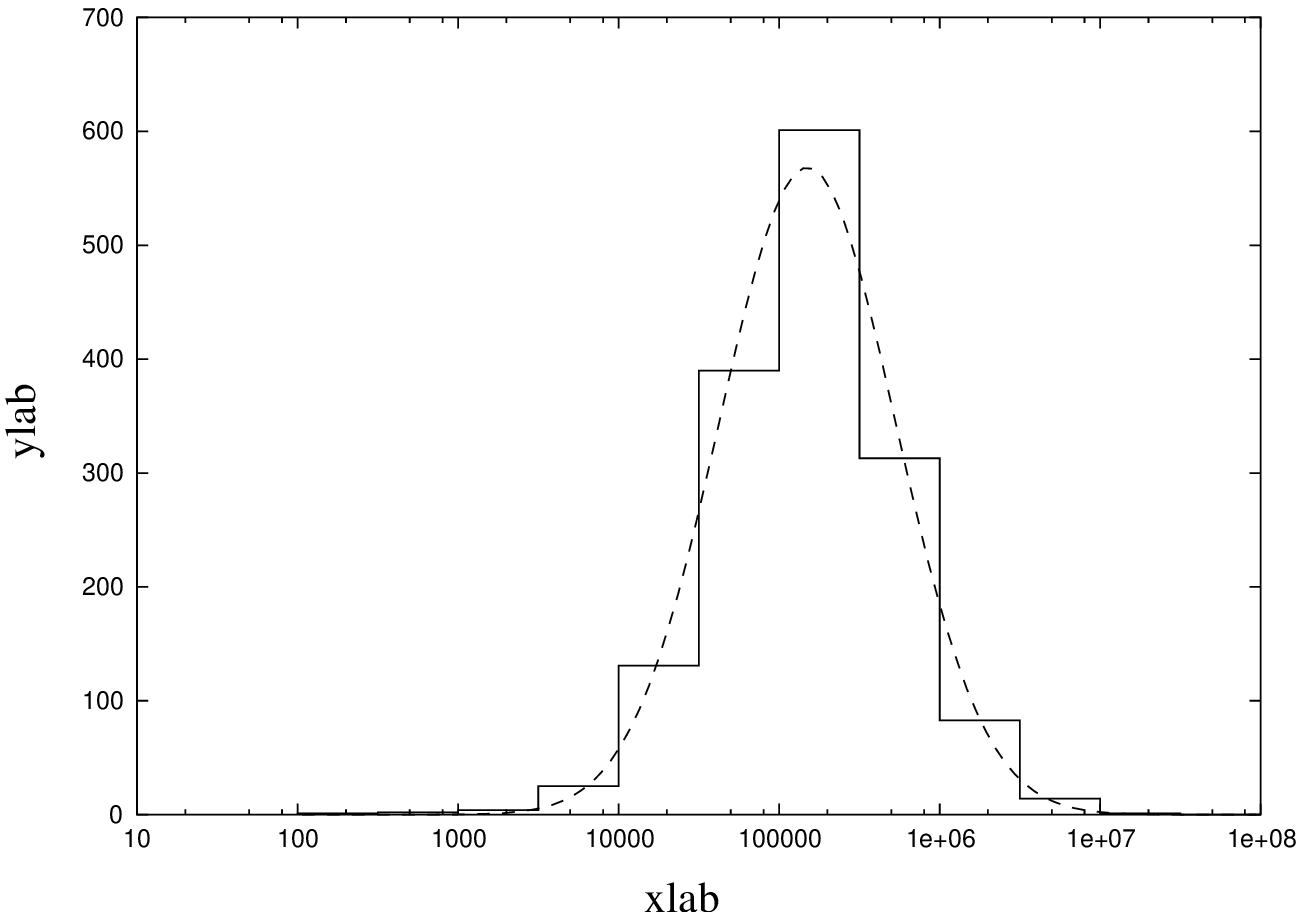}
    \caption{%
    The distribution of the pulse areas with $\tau_{\sigma} \geq 5$ and
    $\tau_{I} \geq 50\%$ and the best lognormal fit to the data.}
\end{figure}


\begin{figure}
    \leavevmode
\resizebox{\columnwidth}{!}{\rotatebox{-90}{\includegraphics{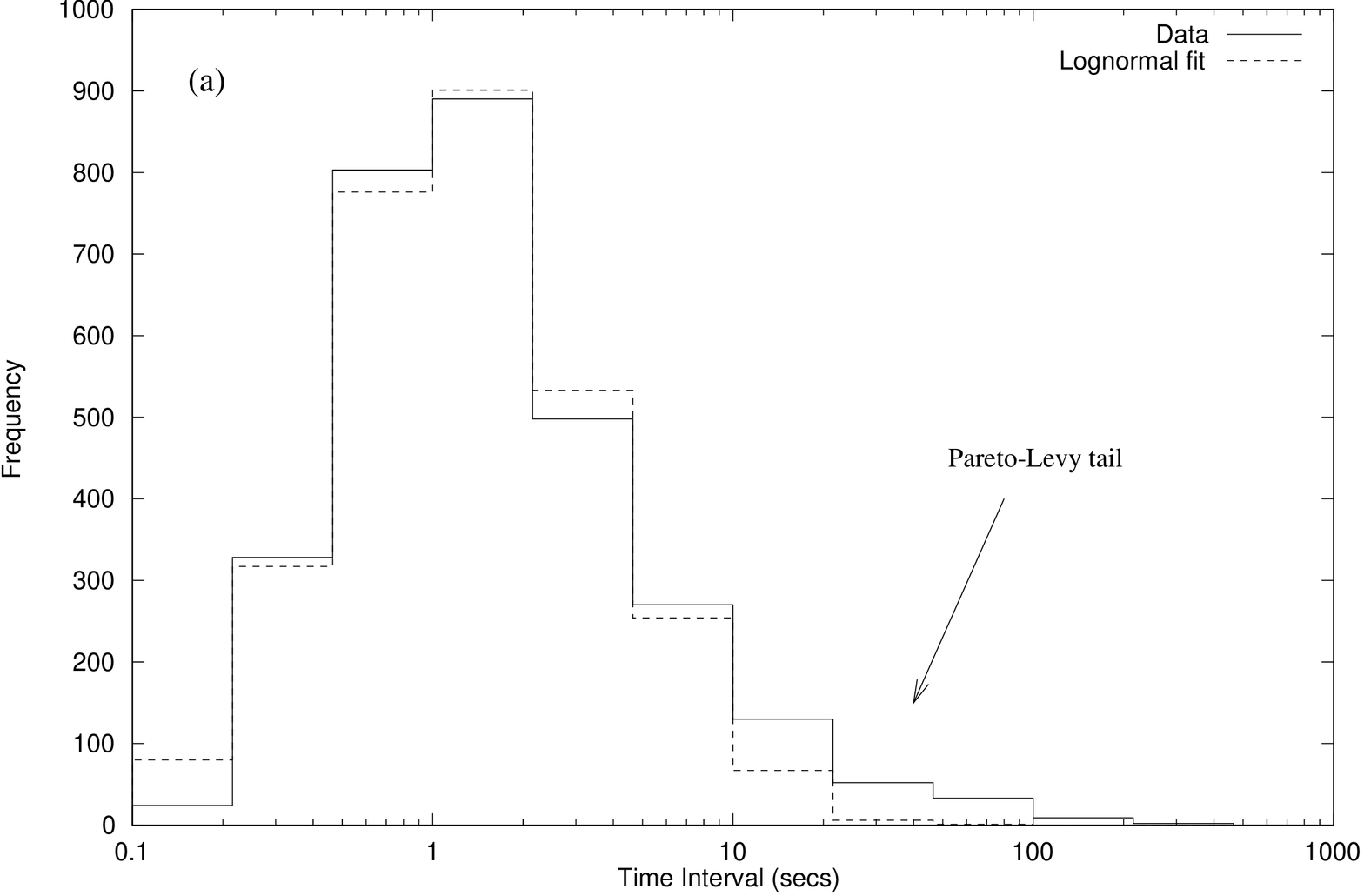}}}

\vspace{1.5em}
    \psfrag{ylabel}[b]{\small Time Interval (sec)}
    \psfrag{xlabel}[b]{\small Cumulative Percentage}
\resizebox{\columnwidth}{!}{\rotatebox{-90}{\includegraphics[width=0.8\columnwidth]{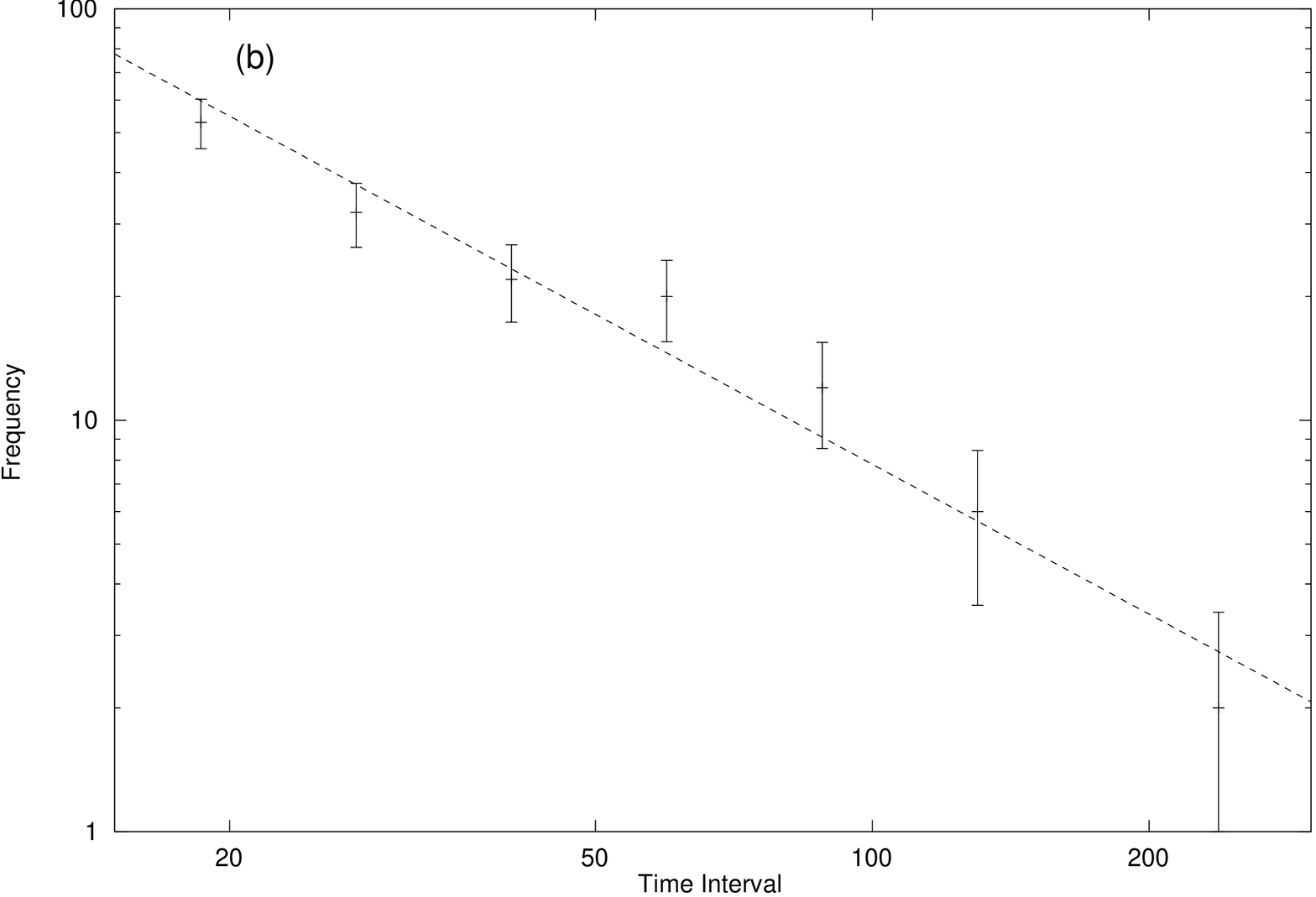}}}
    \caption{%
    (a) The measured distribution of time intervals between all the
    pulses and the best fit lognormal (dashed line) with allowance for the BATSE resolution of 64 ms.
    The excess of time intervals $>$ 15 s is called the Pareto L\'{e}vy tail. (b) The Pareto tail of the time
intervals is well fit by a power law (dashed line) of slope $\sim$
-1.2.}
\end{figure}


\begin{figure}
  \begin{center}
\resizebox{\columnwidth}{!}{\rotatebox{-90}{\includegraphics{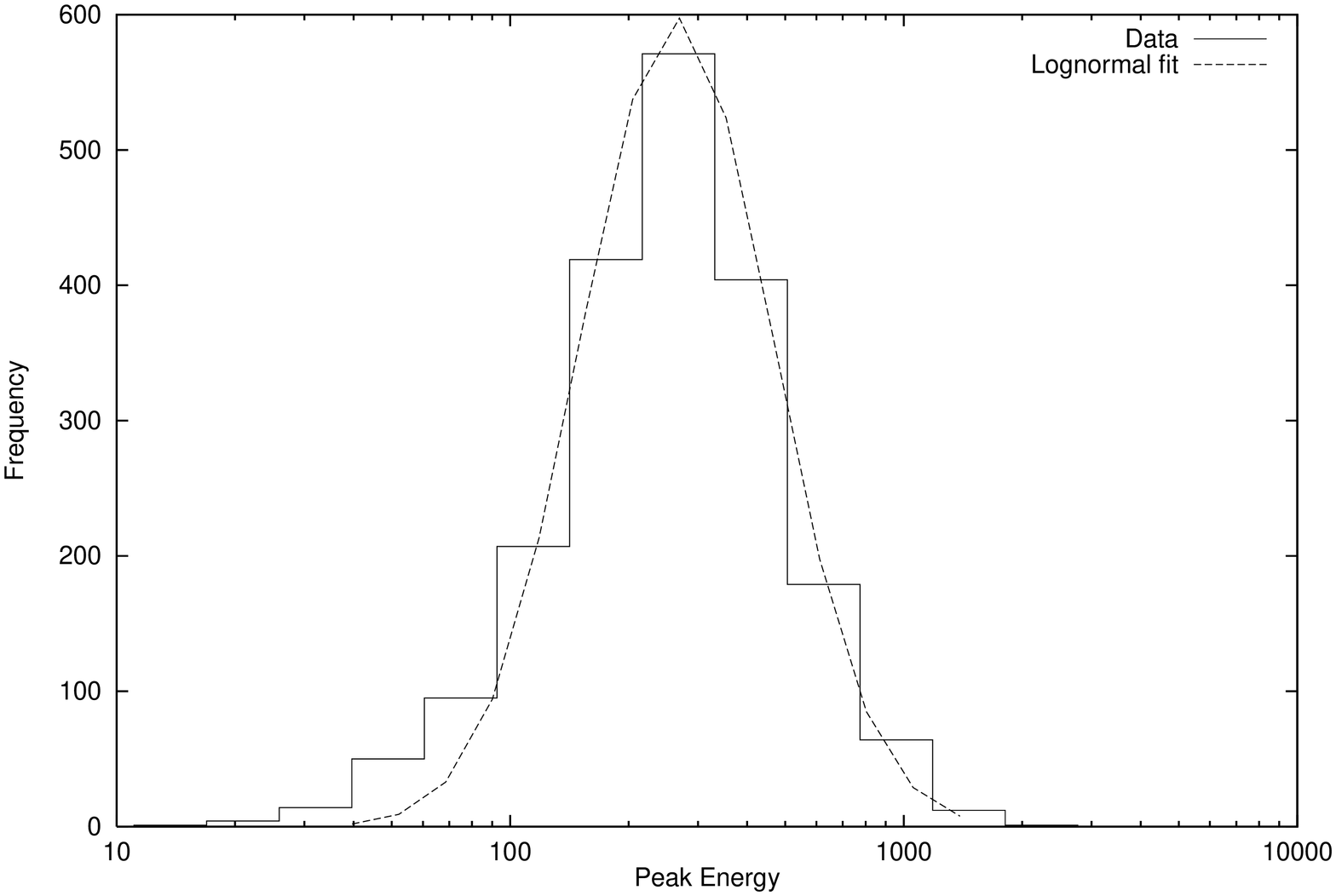}}}
    \caption{%
      The distribution of the peak energies and the best lognormal fit to the data (dashed line).}
  \end{center}
\end{figure}

\subsection{Pulse analysis}
\subsubsection{FWHM of pulses}

The distribution of the FWHM of the pulses from 319 GRBs and the
best lognormal fit are given in Fig 6a.  Fig. 6b shows the same
data plotted as a cumulative percent on logarithmic probability
paper such that a lognormal distribution yields a straight line.
All pulses have $\tau_\sigma \geq 5$ and $\tau_{I} \geq 50\%$.
The distribution is very broad with a maximum at 0.7 s and half
widths at 0.14 s and 3.5 s.  The value of the reduced $\chi^{2}$
is 0.3 showing the data is well fit by the lognormal
distribution.  In obtaining $\chi^{2}$, the part of the fit below
0.1 s was not included because of the distortion caused by the
limited time resolution that is apparent in Fig. 6b.  The value
of the reduced $\chi^{2}$ for the best lognormal fit as a
function of $\tau_{\sigma}$ and $\tau_{I}$ are given in Figs. 6c
and 6d.  The fits are acceptable over most of the range with the
largest departures occurring at the lowest values of
$\tau_{\sigma}$ and $\tau_{I}$ because of the serious effects of
pulse pile up.  The lognormal distribution is a better fit when
the effects of the overlapping pulses are reduced.  The
parameters of the best lognormal fit for $\tau_{\sigma} \geq 5$
and $\tau_{I} \geq 50\%$ are given in Table 2.

\subsubsection{Rise time and fall times of the pulses}

The distribution of the rise times of the pulses and the best
lognormal fit are given in Fig. 7a.  Fig. 7b shows the same data
plotted as a cumulative percent and the large count in the first
bin is due to the 64 ms resolution of BATSE.  The parameters of
the best lognormal fit to the broad distribution are listed in
Table 2.  The value of the reduced $\chi^{2}$ = 1.3 show the data
is compatible with the lognormal distribution.

The distribution of the fall times of the pulses and the best
lognormal fit are given in Fig. 8.  The parameters of the best
lognormal fit to the data are given in Table 2.  The fall times
are also compatible with the lognormal distribution and have a
wider range with a slightly larger mean than the rise times.

\subsubsection{Areas and amplitudes of the pulses}

The distributions of the amplitudes and areas of the pulses with
$\tau_{\sigma} \geq 5 $ and $\tau_{I} \geq 50\%$ from 250 GRBs,
summed over two detectors, and the best lognormal fits are given
in Fig. 9 and Fig. 10.  The distributions are very broad and the
values of the best lognormal fits to the data are listed in Table
2.  The lognormal distribution is compatible with the pulse areas
and the amplitudes.

\begin{table}
\begin{minipage}{\columnwidth}
 \caption{The parameters of the best lognormal fit. The parameters are expressed as natural
 logarithms, with the width of the distributions shown in normal space.}
\setlength{\extrarowheight}{1pt} \addtolength{\tabcolsep}{-1pt}
\begin{tabular}{@{}lcccc@{}}
\hline\hline
Property&$\mu$&$\sigma$&$\chi^{2}$& Width
($\pm$50\%)\\ \hline
 FWHM & -0.36 & 1.37 & 0.3 & 0.14 - 3.5\\
 Rise time & -0.44 & 1.59 & 1.3 & 0.1 - 4.2 \\
 Fall time & -0.07 & 1.59 & 1.5 & 0.14 - 6.1 \\
  Pulse Amp.& 9.0 & 1.12 & 0.3 & 2.2$\times10^{3}$- 30$\times10^{3}$ \\
   Area & 11.9 & 1.2& 1.1& 35$\times 10^{3}$ - 600$\times 10^{3}$ \\
 Time Int.\footnote{See Sect.~\ref{sect:timeintervals}}& 0.21 & 1.03 & -- & 0.37-4.14\\
 Peak Energy\footnote{See Sect.~\ref{sect:peakenergies}}& 5.6 & 0.58 & --&137-535\\
\hline
\end{tabular}
\end{minipage}
\end{table}

\subsubsection{Time intervals between pulses}
\label{sect:timeintervals} The distribution of the time intervals
between the pulses is given in Fig.~11. The peak in the
distribution occurred at about 1.0 s and was truncated at short
time intervals by the 64 ms resolution of the data (Fig.~11a).  A
minimum time interval of 128 ms is required because two maxima
must be separated by at least one time bin. There could be a
large additional excess of pulses with separations below the 128
ms resolution of the data that were not resolved.  Other studies
using different statistical methods from those employed here
\citep{pansm:1999,lbp:2000,lee:2000,spada:2000} have also noted a
deficit of time intervals below one second.  A parent lognormal
distribution of time intervals with parameters similar to the
observed distribution was simulated and the time intervals
between the pulses recorded with an accuracy of 64 ms. The
resulting distribution is given in Fig.~11a and the values of
$\mu$ and $\sigma$ for the parent distribution are given in
Table~2. The measured distribution of time intervals in GRBs is
consistent with the parent lognormal distribution provided a small
(5\%) excess of time intervals longer than 15 seconds is not
included. This excess is clearly visible in Fig.~11a.  The time
intervals greater than 15 s are plotted in Fig.~11b. The data is
well fit by a power law of slope -1.2.

\subsubsection{Peak Energies}
\label{sect:peakenergies} The values of the peak energy, E$_{\rm
peak}$, of a large sample of GRBs are given by \citet{pebs:2000}.
There is an overlap of 77 GRBs with our sample, and the
distribution of the values of E$_{\rm peak}$ for each section of
these bursts are given in Fig.~12. The distribution was well fit
by a lognormal distribution with a small tail noticeable at low
values of E$_{\rm peak}$. The values of $\mu$ and $\sigma$ are
given in Table~2. The distribution of E$_{\rm peak}$ is
noticeably narrower than that of the pulse parameters and spans a
range of about 4 in width.

\subsection{Summary}

The major result of this part of the analysis is that the
distributions of the rise times, fall times, FWHM, pulse
amplitudes, pulse areas and the time intervals between the pulses
are all very similar. The frequency distributions are very broad
and cover about three orders of magnitude and all are compatible
with the lognormal distribution. \citet{lifen:1996} also showed
that the pulse fluences and the time intervals between pulses are
lognormally distributed for individual bursts in a small sample
of bursts with more than 20 pulses. They also scaled the bursts
to the same $\mu$ and $\sigma$ and showed that the summation of
all the peak fluences and time intervals looked lognormal,
although no significance level was given for the result. No
normalisation of pulse properties was applied to the GRBs in this
analysis, because only the brightest 319 bursts which had the
best signal to noise ratio were analysed. From the data available
for those bursts with known redshifts \citep{abh:1999}, there
does not appear to be any dependable standard by which to scale
the bursts because of the broad range of intrinsic luminosities
and their comparatively small range of distances. Therefore, to
avoid introducing further biases, and to use all of the pulse
information available, the data were analysed without scaling.
However, as a test of this process, the data were also scaled and
the same analysis performed on the scaled data, and no
significant differences were found between the two data sets.

\subsection{Correlations between burst and pulse parameters}

It is important to determine how N relates to the other parameters
of the GRB.  In Fig.~13 N is plotted versus burst duration
(T$_{90}$), total fluence and the median value of E$_{\rm peak}$.
Spearman rank order correlation co-efficients $\rho$ and
associated probabilities were obtained for the quantities in
Fig.~13. The values are listed in Table~3 which also includes an
additional range of burst parameters.  The parameter C$_{max}$ is
the maximum value of the peak amplitude in that burst. The high
values of $\rho$ show a strong correlation between N and the total
fluence, T$_{90}$ and E$_{\rm peak}$.

\begin{figure}[htbp]
    \psfrag{xlab}[b]{\small T90(secs)}
    \psfrag{ylab}[b]{\small Number of Pulses}
\includegraphics[width=0.9\columnwidth]{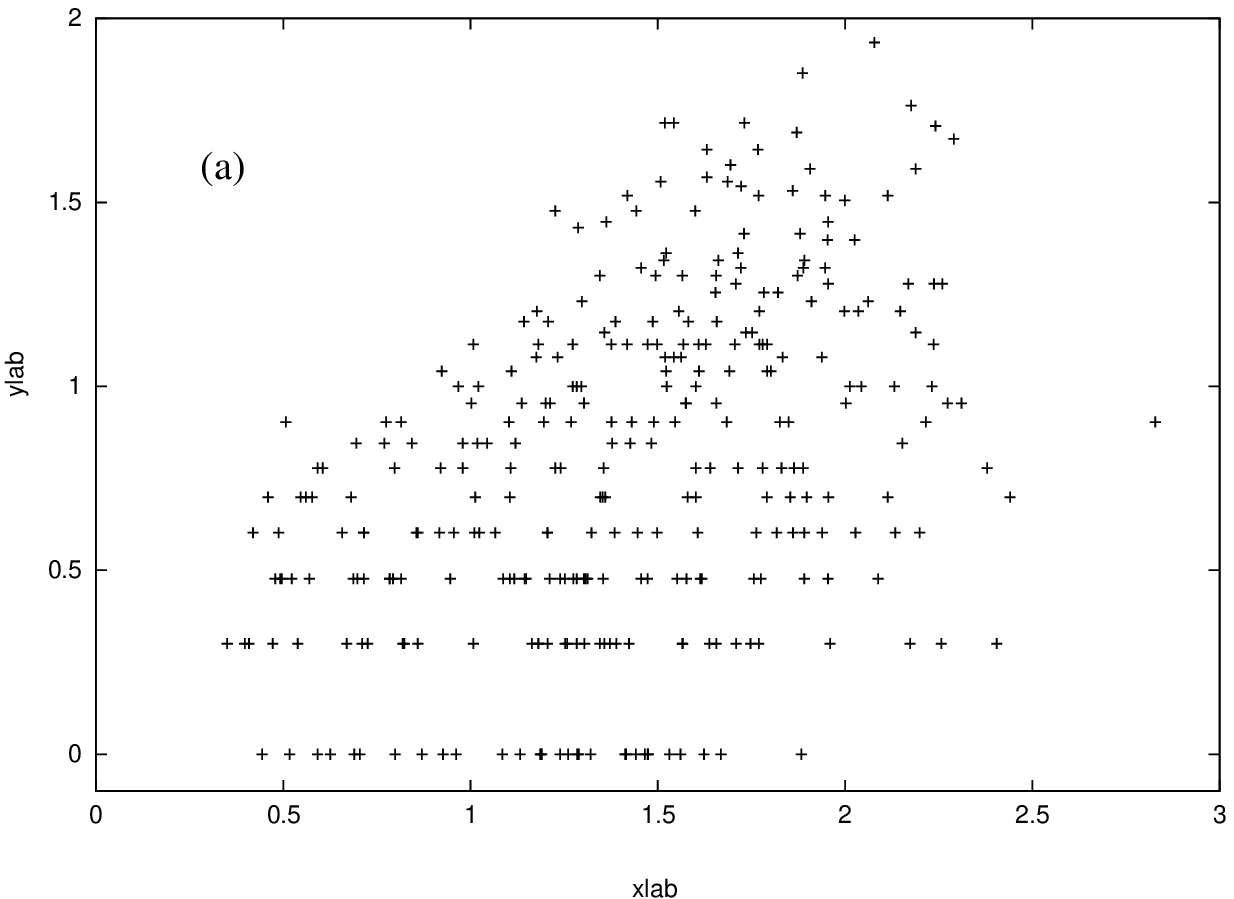}
     \vspace{1.5em}
    \psfrag{xlab}[b]{\small Total Fluence (ergs/cm$^2$)}
    \psfrag{ylab}[b]{\small Number of Pulses}
\includegraphics[width=0.9\columnwidth]{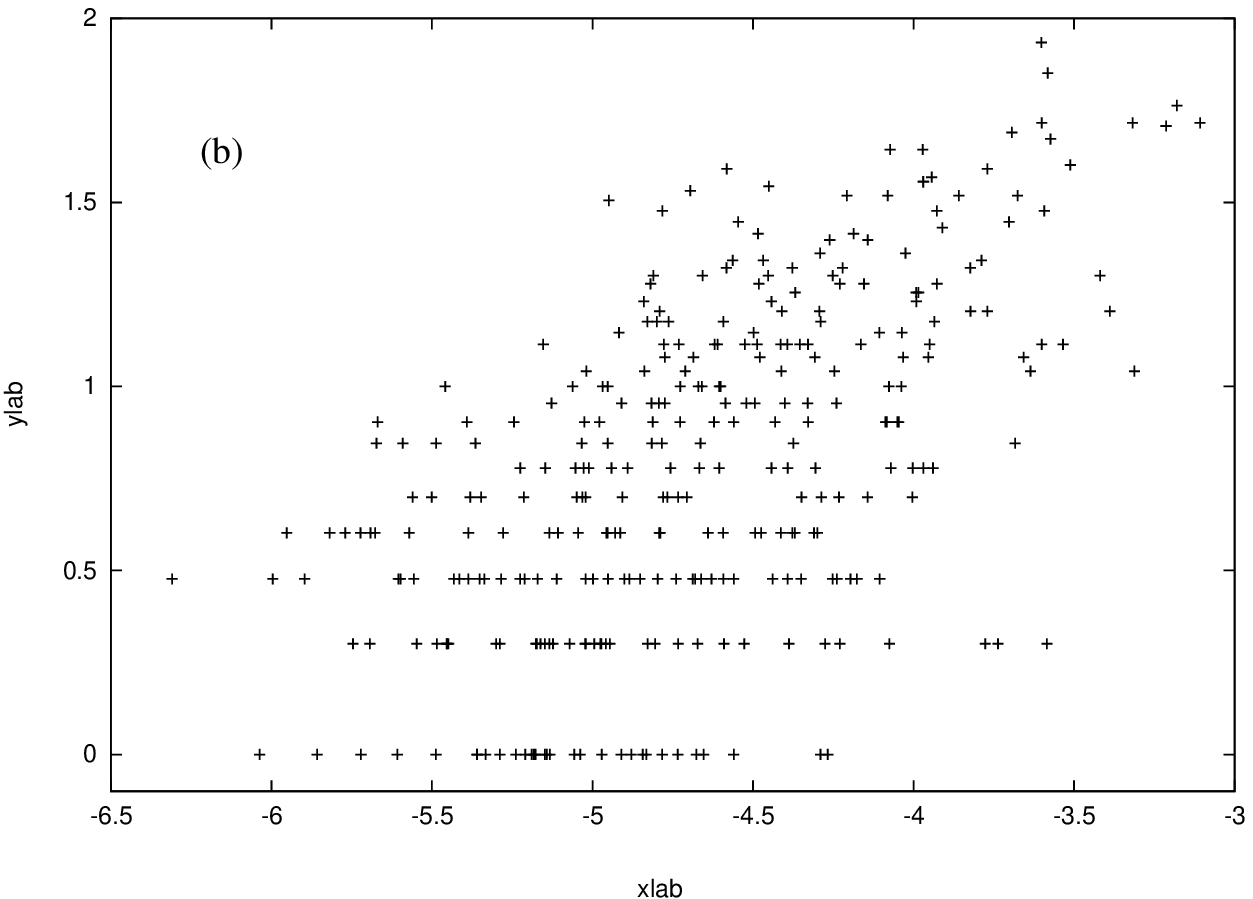}
    \psfrag{xlab}[b]{\tiny Epeak(keV)}
    \psfrag{ylab}[b]{\tiny Number of Pulses}
\resizebox{0.9\columnwidth}{!}{\rotatebox{-90}{\includegraphics[width=0.6\columnwidth]{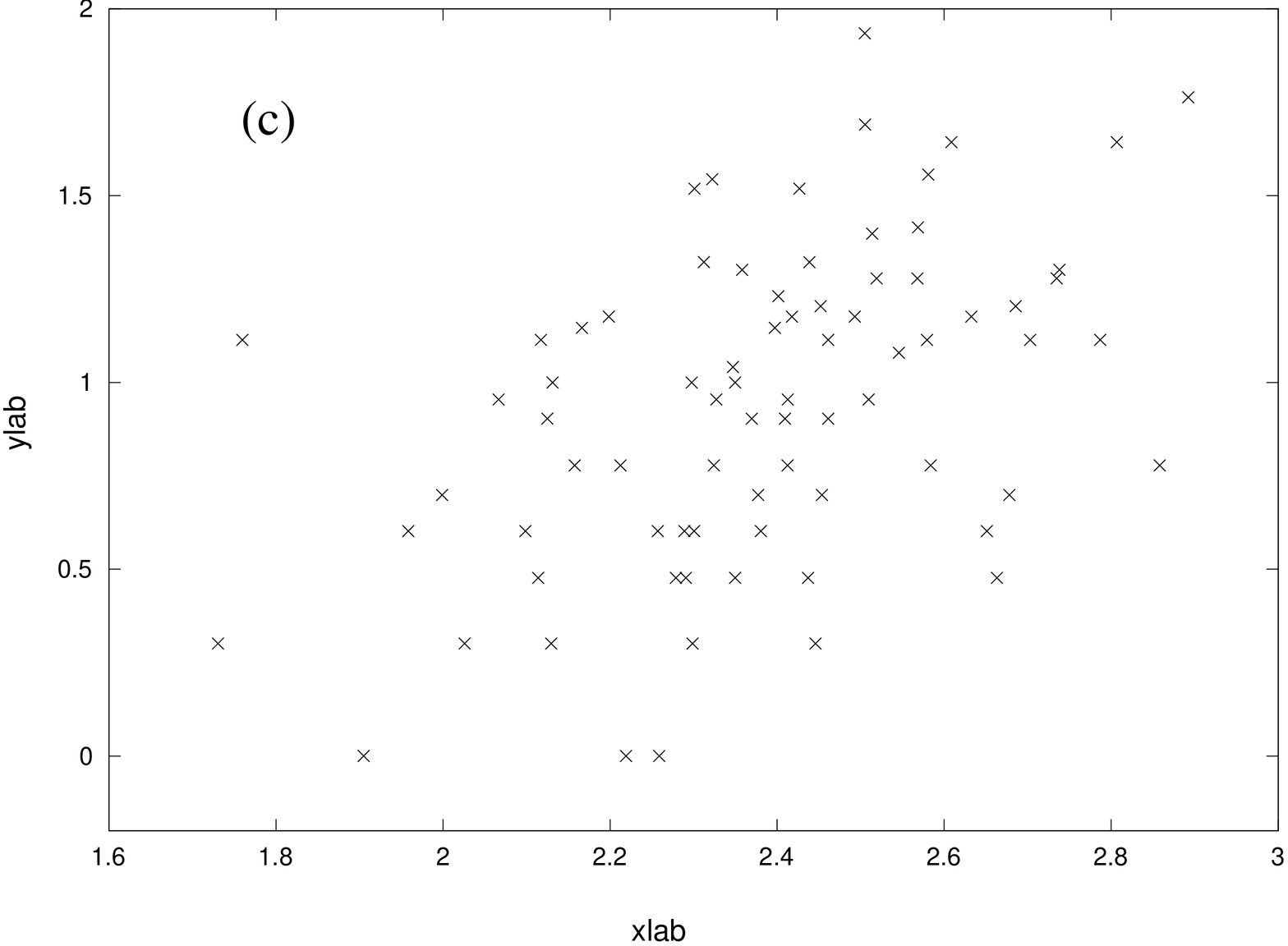}}}
    \caption{
    The number of pulses in a GRB
    as a function of (a) $T_{90}$ (b) total fluence and (c) the
    median value of the peak energy of the GRB.  Quantisation occurs
    in the figures for low values of N.
      }
\end{figure}

\begin{table}
 \caption{Spearman rank order correlation coefficients between a range of
burst parameters}
\setlength{\tabcolsep}{.15cm}
\begin{tabular}{@{}lcr@{}}
\hline\hline
Properties & $\rho$ & Probability \\
\hline
N vs. T$_{90}$  & 0.45 & $5 \times 10^{-17}$\\
N vs. Total Fluence  & 0.58 & $3 \times10^{-30}$\\
N vs. C$_{max}$ & 0.27 & $6 \times 10^{-7}$\\
N vs. Hardness Ratio & 0.29 & $1.2 \times 10^{-7}$\\
N vs. E$_{\rm peak}$ & 0.49 & $7 \times 10^{-6}$ \\
T$_{90}$ vs.\ Total Fluence & 0.52 & $6 \times 10^{-24}$\\
T$_{90}$ vs. C$_{max}$ & -0.08 & 0.16\\
T$_{90}$ vs. Hardness Ratio & 0.11 & 0.05\\
Total Fluence vs. Maximum Pulse  & 0.48 & $1.3 \times10^{-15}$\\
Total Fluence vs. Hardness Ratio (HR) & 0.56 & $3 \times 10^{-27}$\\
C$_{max}$ vs. HR & 0.25 & $4.4\times10^{-6}$\\
\hline
\end{tabular}
\end{table}

\begin{table}
\begin{minipage}{\columnwidth}
\caption{Spearman rank order correlation coefficients between a
range of pulse parameters.} \setlength{\tabcolsep}{.15cm}
\begin{tabular}{@{}lcr@{}}
\hline\hline
Properties & $\rho$ & Probability\\
\hline
Rise Time vs.\ Fall Time & 0.64 &$<10^{-48}$\\
Rise Time vs. FWHM & 0.65 & $<10^{-48}$\\
Rise Time vs. Pulse Area & 0.34 &$2.9\times10^{-34}$\\
Rise Time vs. Pulse Amplitude &-0.27 &$1.9\times10^{-21}$\\
Fall Time vs. FWHM & 0.70 & $< 10^{-48}$\\
Fall Time vs Pulse Area & 0.42 &$< 10^{-48}$\\
Fall Time vs. Pulse Amplitude &-0.22 &$1.4 \times 10^{-14}$\\
FWHM vs. Pulse Area & 0.44 & $< 10^{-48}$\\
FWHM vs. Pulse Amplitude & -0.27 & $2.7 \times 10^{-21}$\\
Pulse Area vs. Pulse Amplitude & 0.63 & $< 10^{-48}$\\
FWHM vs. Time Interval & 0.58 & $10^{-48}$\\
\hline
\end{tabular}
\end{minipage}
\end{table}

The values of $\rho$ are not always uniformly distributed within
each burst category.  T$_{90}$ versus fluence is much better
correlated for category N than either O or P.  N versus C$_{max}$
and N versus HR are better correlated for category P than either N
or O.

The Spearman rank order correlation coefficients and probabilities
were evaluated for isolated pulses with the range of pulse
parameters given in Table~4.  The pulse parameters are strongly
correlated with each other.  The pulse amplitude is negatively
correlated with the pulse rise and fall times and FWHM.  In
general the correlations are stronger for categories M and N than
either O or P. The only significant exception to this trend is
the pulse amplitude versus area which also has the highest values
of $\rho$ for categories O and P.

\subsection{Correlations between the time intervals between
pulses and pulse amplitudes}

Spearman rank order correlation coefficients and probabilities
were evaluated for the time intervals between pulses ($\Delta$T)
with $\tau_{\sigma} >$ 5.  The results are presented in Table 5
for two cases (1) the time intervals were not normalised and (2)
the time intervals were normalised to T$_{90}$.  There is a good
correlation between the time intervals in both cases that
declines slowly with increase in the number of time intervals.
The largest values of $\rho$ occured in category N.

The Spearman correlation coefficients were also evaluated between
pulse amplitudes and the results are given in Table~6 for two
cases (1) the amplitudes were not normalised and (2) normalised
to the largest amplitude pulse in the burst.  The normalised
pulse amplitudes are less strongly correlated over many pulses
than the time intervals. These results were obtained for all
pulses with $\tau_\sigma\geq5$ and without satisfying any
selection based on pulse isolation. The role of pulse pile-up has
yet to be investigated.

\begin{table}
\begin{minipage}{\columnwidth}
 \caption{Spearman rank order correlation coefficients $\rho$
for time intervals between pulses.  The two values for $\rho$ and
the probability are for unnormalised/normalised time intervals.}
 \setlength{\tabcolsep}{.15cm}
\begin{tabular}{@{}lccr@{}}
\hline\hline Number of& Total & & \\
Intervals & Number& $\rho$ & Probability\\
\hline
1 & 2751 & 0.42/0.56 & $< 10^{-48}$\\
2 & 2499 & 0.34/0.48 & $< 10^{-48}$\\
5 & 1929 & 0.24/0.37 & $5\times10^{-26}/<10^{-48}$\\
10 & 1395 & 0.20/0.29 & $3\times10^{-13}/6\times10^{-27}$\\
15 & 890 & 0.16/0.25 & $3\times10^{-6}/4\times10^{-14}$\\
20 & 634 & 0.10/0.23 & $8\times10^{-3}/3\times10^{-9}$\\
25 & 459 & 0.08/0.22 & $7\times10^{-2}/1\times10^{-6}$\\
30 & 322 & 0.03/014 & $3\times10^{-2}/10^{-2}$\\
\hline
\end{tabular}
\end{minipage}
\end{table}

\begin{table}
\begin{minipage}{\columnwidth}
\caption{Spearman rank order correlation coefficients $\rho$ and
associated probabilities for the pulse amplitudes. The two values
for $\rho$ and the probability are for unnormalised/normalised
amplitudes.} \setlength{\tabcolsep}{.15cm}
\begin{tabular}{@{}lccr@{}}
\hline\hline Number of & Total & & \\
Amplitudes & Number& $\rho$ & Probability \\
\hline
1 & 3039 & 0.72/0.57 &  $< 10^{-48}$\\
3 & 2499 & 0.55/0.32 &  $< 10^{-48}$\\
5 & 2098 & 0.52/0.24 &  $< 10^{-48}/ 3\times10^{-29}$\\
7 & 1777 & 0.48/0.15 & $< 10^{-48}/ 6\times10^{-11}$\\
9 & 1510 & 0.43/0.08 & $< 10^{-48}/2 \times 10^{-3}$\\
10 & 1395 & 0.44/0.08 & $< 10^{-48}/3 \times 10^{-2}$\\
\hline
\end{tabular}
\end{minipage}
\end{table}

\subsection{The properties of the pulses as a function of, N, the number of pulses
in the GRB}

It was  noticed early in this analysis that pulse properties
depended strongly on N \citep{quilligan:2000}.  The median value
of the isolated pulse timing parameters were determined for all
GRBs with the same value of N.  The median values of the rise
time, fall time, FWHM and time interval between pulses are plotted
versus N in Fig.~14(a-d). The largest value usually occurred for
N = 1 or 2 and subsequently declined significantly as N
increased.  There are some values that are well removed from the
general trend but they usually have a small number of pulses. The
median values of the area and amplitude for isolated pulses are
given in Fig.~15 (a-b). The trend is quite different from
Fig.~14.  The amplitude is reasonably constant up to N $\sim$ 35
with a clear increase for higher values of N. There is a similar
but weaker trend for the pulse area which has the largest value
at N = 1.

The properties of the four categories of GRBs are summarised in
Table~7. The median values of the pulse timing parameters all
decrease by at least a factor of four from category M to P. In
contrast the median values of T$_{90}$, total fluence, hardness
ratio and maximum pulse amplitude all increase significantly. The
median variability, is defined as the number of pulses $\geq$
5\,$\sigma$ divided by the time the emission is $\geq$ 5 $\sigma$,
also increases from category M to P.

\begin{table*}
 \caption{The properties of the four categories of pulses in GRBs.
 The last five entries are for the 250 GRBs that were summed over two detectors.}
 \setlength{\tabcolsep}{0.6cm}
\begin{tabular}{@{}lcccr@{}}
\hline\hline
GRB Category  & M  & N & O & P\\
\hline
Number of Pulses per GRB  & 1-2 & 3-12 & 13-24 & 25+\\
Number of GRBs  & 67& 162 & 56  & 34\\
Total number of pulses  & 103& 981  & 933 & 1341   \\
Number of isolated pulses at 50\% level & 83 & 522 & 476 & 494\\
Median T$_{90}$ (sec) & 18.1& 20.4  &  45.7     & 58.7    \\
Median Total Fluence (ergs/cm$^2$)& 8.8$\times 10^{-6}$  & 1.7$\times 10^{-5}$ &4.2$\times 10^{-5}$ &1.2$\times 10^{-4}$\\
Median hardness ratio (Chan $\frac{4+3}{2+1}$)& 3.4& 4.1 & 6.5 & 8.3\\
Median $C_{max}$ (ph/cm$^2$/sec)& 4.7 & 6.1 & 9.3 & 12.6\\
Median Variability & 0.09 & 0.29 & 0.39 & 0.53\\
Median Rise Time (sec) & 1.7 & 0.8  & 0.64 & 0.45\\
Median Fall Time (sec)& 5.2 & 1.5 & 1.0 & 0.7\\
Median FWHM (sec) & 1.8 & 0.7 & 0.64 & 0.45\\
Median Time Interval (sec)& 4.8 & 1.9 & 1.5& 1.0\\
\hline
Number of GRBs  & 55 & 130 & 38  & 27\\
Total number of pulses  & 87& 778  & 648 & 1081   \\
Number of isolated pulses at 50\% level & 70 & 416 & 319 & 404\\
Median Pulse Amplitude (Iso. pulses) (counts)& 13$\times 10^{3}$ & 7.3$\times 10^{3}$ & 7.3$\times 10^{3}$ & 9.5$\times 10^{3}$\\
Median Area (counts)& 560$\times 10^{3}$ & 180$\times 10^{3}$ & 140$\times 10^{3}$ & 140$\times 10^{3}$\\
\hline
\end{tabular}
\end{table*}

\begin{figure}
  \begin{center}
\includegraphics[height=0.95\textheight]{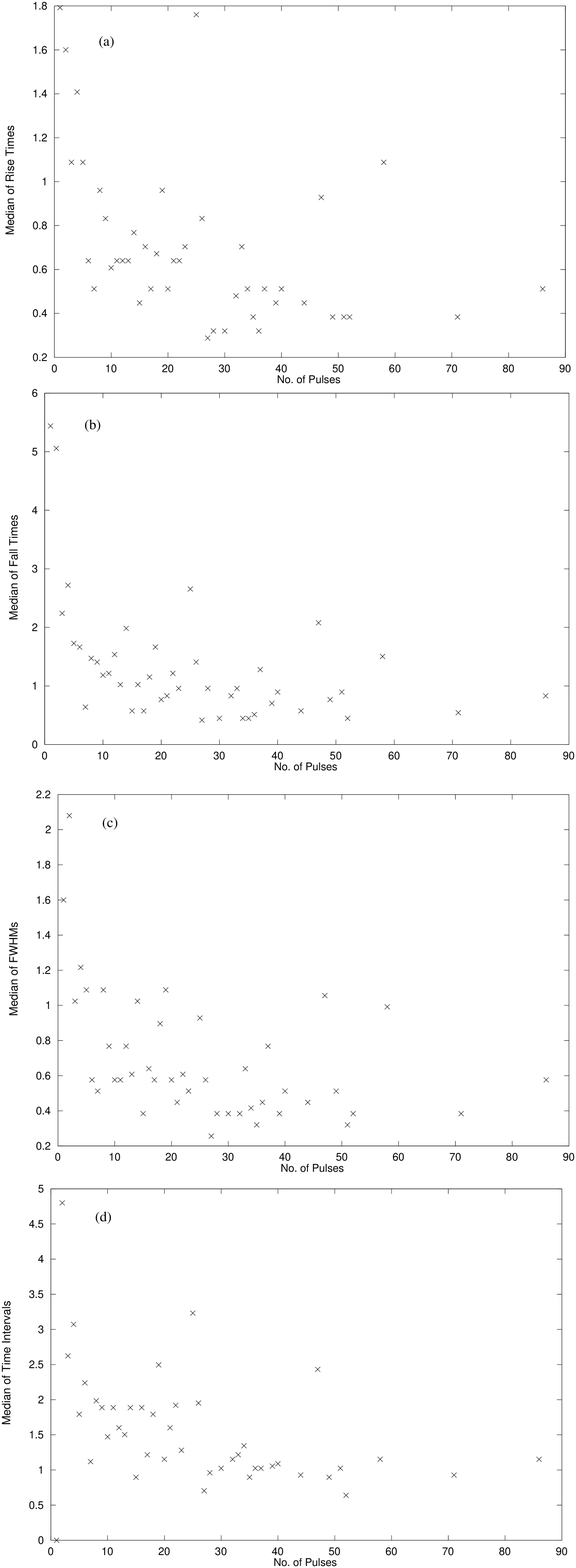}
    \caption{The median values of (a) rise time, (b) fall time,
    (c) FWHM and (d) the time intervals between pulses versus the
    number of pulses}
  \end{center}
\end{figure}

\begin{figure}
  \begin{center}
\resizebox{0.95\columnwidth}{!}{\rotatebox{-90}{\includegraphics[width=0.5\columnwidth]{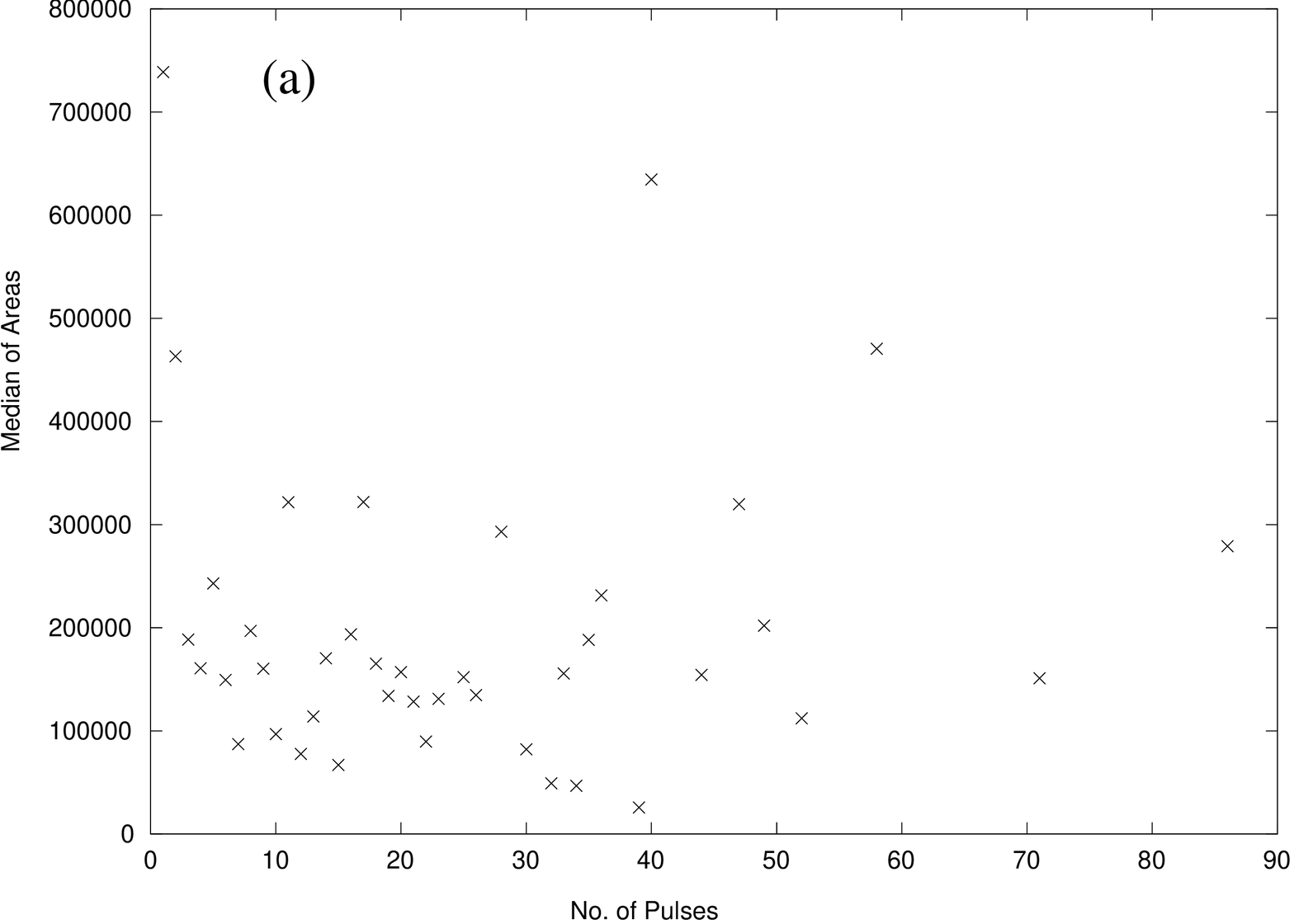}}}
\resizebox{0.95\columnwidth}{!}{\rotatebox{-90}{\includegraphics[width=0.5\columnwidth]{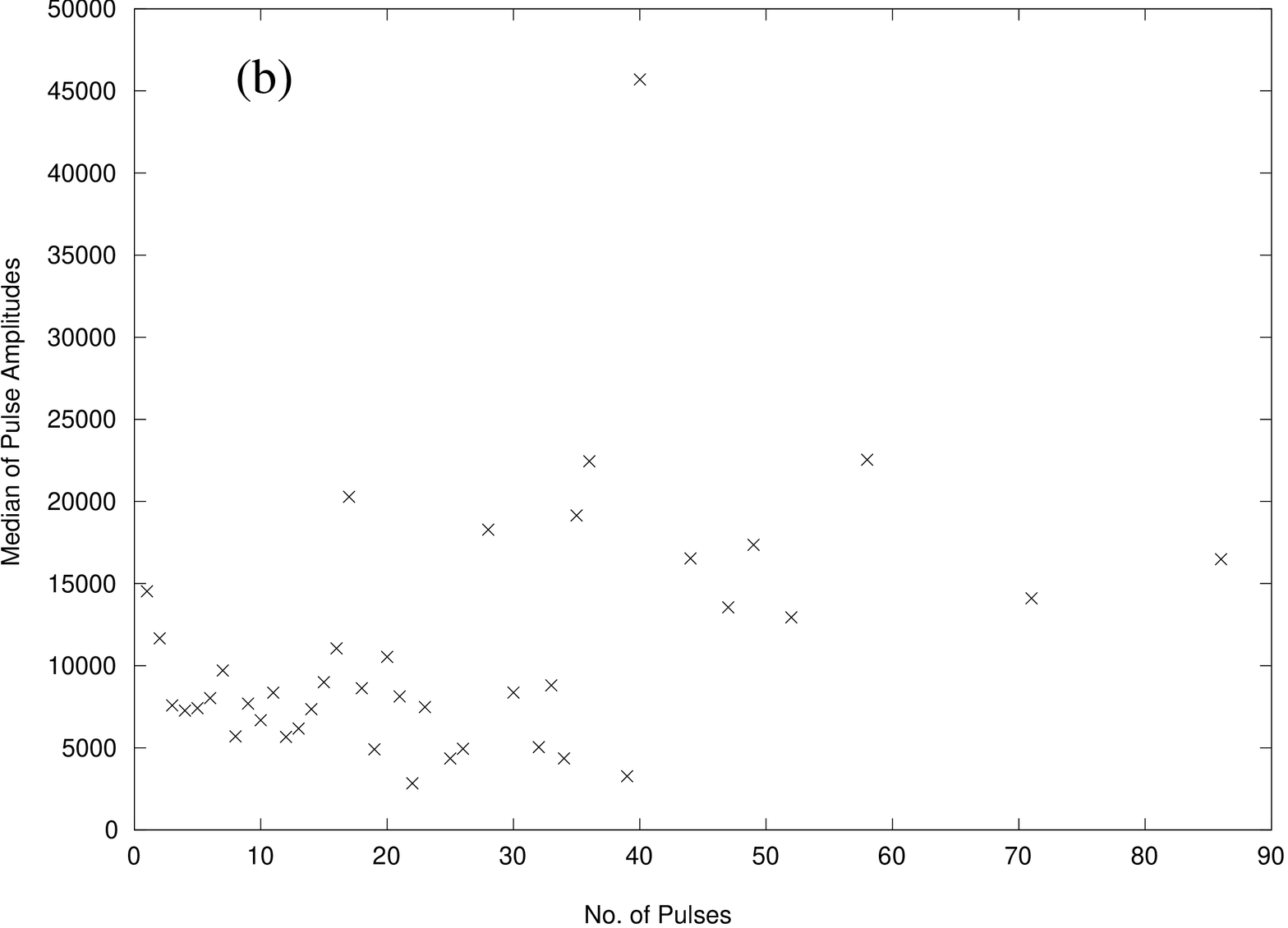}}}
    \caption{The median values of (a) pulse area and (b) pulse
    amplitudes versus the number of pulses.}
  \end{center}
\end{figure}

\subsection{First Half/Second Half analysis}

\begin{table*}
 \caption{Summary of the first half/second half analysis of the three categories of the GRBs.}
\setlength{\tabcolsep}{0.5cm}
\begin{tabular}{@{}lrrr@{}}
\hline\hline
GRB Category   & N & O & P\\
\hline
Number of Pulses per GRB  & 3-12 & 13-24 & 25+\\
Total Number of Pulses (1st/2nd half) & 404/415 & 384/493 & 679/628\\
Total Number of Isolated Pulses (50\%) (1st/2nd half) & 193/220 & 182/261 & 203/271\\
Total Number of Isolated Pulses (75\%) (1st/2nd half) & 79/126 & 77/101 & 71/83\\[8pt]
Median Rise Time (1st/2nd half)& 1.02/0.70 & 0.77/0.58 & 0.51/0.45\\
K-S Statistic/Probability & 0.13/5\% & 0.12/9\%  &0.11/11\%\\[8pt]
Median Fall Time (1st/2nd half)& 1.09/1.89 & 0.90/1.22 & 0.70/0.70\\
K-S Statistic/Probability & 0.20/.02\% & 0.15/2\% & 0.11/10\%\\[8pt]
Median Asymmetry Ratio (1st/2nd half) & 1.0/0.4 & 0.75/0.49 & 0.75/0.54\\
K-S Statistic/Probability & 0.27/$3\times 10^{-5}\% $ & .17/0.3\% & 0.14/1.5\%\\[8pt]
Median FWHM (1st/2nd half)& 0.77/0.67&  0.64/0.64 & 0.51/0.45\\
K-S Statistic/Probability & 0.08/55\%& 0.08/52\% & 0.11/11\%\\[8pt]
Median Time Interval (1st/2nd half)& 1.60/2.10 & 1.41/1.54 & 1.02/1.02\\
K-S Statistic/Probability & 0.13/2\% & 0.06/35\% & 0.05/30\%\\[8pt]
Median Pulse Amplitude  ($\times 10^{3}$)(1st/2nd half) & 6.2/5.4 & 7.3/6.8 & 18.1/11.2\\
K-S Statistic/Probability & 0.09/41\% & 0.08/52\% & 0.09/25\%\\[8pt]
Median Area ($\times 10^{3}$)(1st/2nd half) & 151/127 & 133/118 & 152/130\\
K-S Statistic/Probability & 0.08/50\% & 0.07/59\% & 0.14/2\%\\
\hline
\end{tabular}
\end{table*}

 To study the evolution of the time profile as the GRB
progresses, each GRB was divided into two, to include the pulses
that occur before and after the strongest pulse in the burst.
Only GRBs with more than two pulses are included, resulting in a
reduced total sample of 252. The first half (pre-main pulse) of
the GRB was compared with the second half (post-main pulse). The
bursts are also sub-divided into three categories.  A summary of
the properties of the GRBs used is given in Table~8.

The first half/second half analysis was performed on the three
timing parameters of the pulses, time intervals between the
pulses, amplitude, area and the pulse asymmetry ratio which is
defined as the ratio of the pulse rise time to the pulse fall
time. The median values of the distributions in the three
categories in the first half and second half analysis are given in
Table~8 along with the results of the Kolmogorov-Smirnov (KS)
tests.  The KS probability is a measure of whether the two
distributions (first/second half) are drawn from the same parent
distribution.

The first result is that the median values of the timing
parameters of the pulses and the time intervals between the pulses
all decrease by an average of 1.8, from the category N to P
including the first half and second half of the GRBs.  In the case
of the rise times, the median values of the distributions decrease
from 1.02 to 0.51 in the first half and 0.70 to 0.45 in the second
half. The trend in the median value of the pulse amplitude is in
the opposite direction with larger pulses in category P than
either O or N.

In the first/second half analysis there is a trend in the three
categories for the median rise time to be slower in the first half
of the burst (1.02 versus 0.70 for category N).  The difference
could be caused by an additional clearing out effect at the start
of the GRB. There is also a clear indication at the 0.02\% level
that the pulse fall time is faster in the first half than the
second for category N (1.09 versus 1.89 for category N) and this
effect weakens for categories O and P.  The median values of the
pulse asymmetry ratio also show the most significant differences
for category N where the median values are 1.0 and 0.4 for the
first and second halves.  The KS test gives good agreement
between the first half and second half for the FWHM, time
intervals between pulses, the pulse amplitudes and areas. The
median values of the pulse amplitude and areas are however larger
in the first half than the second half for the three categories
of GRBs.

\section{Discussion}

\subsection{Pulse shapes in GRBs}

A complete study of the BATSE time profiles of the brightest 319
GRBs has been presented. The statistical analysis of the data
reveal the ubiquitous nature of the lognormal distribution in GRB
time profiles (Figs. 6-10). The means and variances of the best
fit lognormal functions are given in Table~2.  The results
presented in Figs. 6-10 are for the isolated pulses from the four
GRB categories. The data in Table~7 and the first half/second
analysis show that the median values of the pulse properties vary
with N.  Lognormal distributions also apply to the spectral
properties of GRBs.  The BATSE spectroscopic detectors revealed
that the break energies in GRB spectra are compatible with a
lognormal distribution \citep{pebs:2000}.  The FWHM of the
distributions that describe the pulses are in the range 14-45
(Table~2) whereas the value for E$_{\rm peak}$ is only $\approx$ 4
and extends from 137 keV to 535 keV. The unexpected narrowness of
the E$_{\rm peak}$ distribution is a major problem in GRBs
\citep{brain:2000}. Furthermore, spectral fitting of 41 pulses in
26 GRBs showed that the spectral hardness parameter E$_{\rm peak}$
decays linearly with energy fluence and that the distribution of
the decay constants is roughly lognormal \citep{crider:1999}.

The Spearman rank order correlation coefficients were obtained
for a range of burst parameters (Table~3).  There is good
agreement with the results of Lee et al. (2000a) who used a
different method. N is strongly correlated with total fluence,
T$_{90}$, HR and E$_{\rm peak}$ (Fig.~13).  N is an important
quantity in determining GRB properties and provides the link
between total fluence and duration since both increase with N.
The correlation coefficients between the parameters that describe
the pulses are given in Table~4. The rise and fall times, FWHM
and area of the pulses are highly correlated. There is a high
probability that a pulse with a fast rise time will also have a
fast fall time and a short FWHM. The pulse amplitude is
negatively correlated with the rise and fall times and FWHM. The
anticorrelation between the pulse amplitude and pulse width has
been observed in other studies \citep{lbp:2000,feniram:2001}. The
pulse width is also a function of energy and varies as
E$^{-0.45}$ \citep{fenimore:1995} and this effect has been
attributed to synchrotron radiation \citep{piran:1999}. The FWHM
is strongly correlated with the preceding and subsequent time
interval between pulses (Table 4). This result is in agreement
with Nakar \& Piran (2001) and the prediction of the internal
shock model.  A further comparison of the pulse timing
parameters, the energy dependence and the spectral lag may reveal
further interesting constraints on the emission process
\citep{norris:2001,daigne:2001}.  The unique properties of the
pulses in GRBs have been summarised by McBreen et al.,(2002).

The lognormal distribution arises from a statistical process whose
result depends on a product of probabilities arising from a
combination of independent events.  It therefore identifies the
statistical process but not the combination of events that lead to
the formation of the pulse shape and the peak energy.  In the
internal shock model the main factors contributing to the pulse
shape include \citep{reemes:1994,piran:1999}.
\begin{enumerate}
\item the Lorentz factors, masses and thickness of the
interacting shells
\item the distance from the central engine and curvature where
the collisions occur
\item the energy conversion in the shock, the nature of the
magnetic field and particle acceleration process
\item the synchrotron radiation mechanism possibly modified by
self-absorption and pair production
\item the time scale for energy loss by the particles.
\end{enumerate}
The resulting properties of the pulses in GRBs depend on a
combination of many events and hence it is not surprising that the
lognormal distribution gives an elegant description of all their
properties.  Random multiplicative processes abound in a variety
of natural phenomena and a good example is the statistical
description of the strokes in flashes of lightning. Almost all
the properties of the strokes in the flashes and the flashes
themselves are well described by lognormal distributions
\citep{uman:1987,mhlm:1994}.

\subsection{Time intervals between pulses in GRBs}

It is often found that distributions that seem to be lognormal
over a wide range change to an inverse power law distribution for
the last few percent.  An amplification model has been used to
characterise the transition from a lognormal distribution to a
power law that is often called a Pareto-L\'{e}vy tail
\citep{montshles:1982}. The distribution of time intervals conform
to the lognormal distribution over most of the range with the
exception of about 5\% of the time intervals longer than about
15\,s (Fig.~11).  The Pareto-L\'{e}vy tail of time intervals have
an amplification process that is not available to most time
intervals.

The origin of the nonrandom distribution of time intervals between
pulses is an important clue to the GRB process. In the internal
shock model there is almost a one to one correspondence between
the emission of shells and pulses resulting from the collisions of
shells \citep{kps:1997}.  Hence the time intervals between pulses
is an almost direct measure of the activity of the central engine.
The temporal behaviour of soft gamma-ray repeaters and young
pulsars provide additional context in which to view the results of
GRB time profiles. The time intervals between about 30
microglitches in the Vela Pulsar are consistent with a lognormal
distribution with a mean of 50 days
\citep{hmrs:1994,hurley:1995,cordk:1988}. The amount of energy
involved in the microglitches is about $10^{38}$ ergs. The
macroglitches in the Vela Pulsar are about a thousand times more
powerful but occur too infrequently to determine the distribution
of time intervals but they have a wide range and do not seem
inconsistent with the lognormal distribution.  More energetic
outbursts have been recorded from SGR sources. The two most
energetic events released about $5 \times 10^{44}$ ergs in
$\gamma$-rays from SGR 0526-66 on 5 March 1979 \citep{mgi:1979}
and about $10^{43}$ ergs from SGR 1900+14 on 27 August 1998
\citep{hcmb:1999}. The SGR sources also generate a large number of
smaller outbursts, and it has been shown that the time intervals
between outbursts are distributed lognormally
\citep{hmrs:1994,GWK:2000}. Hence lognormally distributed time
intervals between outbursts and glitches are characteristic
features of SGR sources and neutron star microglitches. It is
widely accepted that these sources are rotating neutron stars with
high magnetic fields.  It is not unreasonable to argue that the
coupled effects of rapid rotation and intense magnetic fields
\citep{kluzrud:1998} are also involved in powering GRBs since the
time intervals between pulses are also consistent with a
lognormal distribution.

The possibility that a rapidly rotating neutron star with a
surface magnetic field of $\sim 10^{15}$ Gauss could power a GRB
has been suggested \citep{usov:1992}.  Once formed such a neutron
star could lose its rotational energy catastrophically on a time
scale of seconds. The rotation of the star decelerates because of
the applied torques. Powerful transient fields may also occur in
the merger of two neutron stars or a neutron star and a black
hole.  The energy stored in differential rotation of the collapsed
object would be released in sub-bursts as toroidal magnetic fields
are repeatedly wound up to \(\sim 10^{17}\) Gauss
\citep{kluzrud:1998}.  The emergence of a toroid is accompanied by
huge spin down torques, the reconnection of new surface magnetic
fields and rapid release of a sub-burst of energy of about
10$^{51}$ ergs.  The release of rotational energy in repeated
sub-bursts could power the GRB.  If the differentially rotating
compact object forms a torus about a spinning black hole either in
a merger or core collapse of a massive star, energy can be
extracted by the magnetic field that threads the torus and the
black hole \citep{mes:2001}.  As the torus builds up and ejects
its magnetic toroids, the differential rotation of the torus
could be maintained by the spin of the black hole.The models of
GRBs with the coupled effects of rapid rotation and ultra intense
magnetic fields are particularly attractive because the time
intervals between pulses in GRBs are distributed lognormally and
follow the pattern observed in non-catastrophic events in SGRs
and pulsars.

The time intervals between the pulses are correlated with each
other and the correlation decreases slowly with increase in the
number of time intervals (Table~5). This effect had previously
been observed in a small sample of GRBs \citep{np:2001} and
attributed to the internal shock model. In addition the pulse
amplitudes are also correlated with each other and this effect
decreases more rapidly than the time intervals between pulses
(Table~6). Similar correlations have been found between the pulse
amplitudes and also time intervals between pulses in short GRBs
\citep{sheila:2001}.  In the internal shock model, these
correlations originate in the central engine and provide strong
constraints on any viable model of GRBs.

GRB models leave open many possibilities to account for the
Pareto-L\'{e}vy tail of long time intervals. The excess of long
time intervals have been noted in other studies
\citep{rm:2001,np:2001}. The properties of the GRBs with long time
intervals will be covered in a separate publication.

\subsection{Numbers of pulses and jets in GRBs}

A detailed comparison has been made between the distributions of
the properties of the pulses in the first half and second half for
 three categories of GRBs.  There are no statistically significant
differences between the median values of the time intervals
between pulses, pulse amplitude, areas and FWHM in the first half
and second half of GRBs (Table~8). There are two trends in the
pulse rise times and fall times that should be noted: 1)\, the
median rise time is slower in the first half for the three
categories of bursts (Table~8) and 2)\,the median fall time is
faster in the first half for categories N and O.  The combination
of slower rise times and faster fall times gives a pulse
asymmetry ratio with a significant difference between the first
half and second half for category N and at a reduced significance
level for O and P (Table~8). The effect could be caused by a
clearing out process such as additional baryon loading or Compton
drag in the first half of the bursts with small number of
pulses.  These results are also compatible with the constancy of
the pulse widths observed by \citet{ramfen:2000} using a peak
aligned profile method on a small sample of GRBs with more than
20 pulses. In the internal shock model, the rapid variability in
GRB time profiles is due to emission from multiple shocks in a
relativistic wind \citep{piran:1999,pansm:1999,downes:2001}. The
temporal position of the pulse is unconnected to the collision
parameters and in this way the little or no evolution of the
pulses in GRBs can be explained \citep{feniram:2001}.  The rise
time and fall time may be determined by the hydrostatic time
$\approx$d/c and the angular spreading time $\approx$D/c, where d
and D are the width and separation of the shells
\citep{kobay:2001}. The observed evolution of the pulses requires
the shells to be narrower and farther apart later in the GRB.
This prediction is in agreement with the data because the time
intervals between pulses are longer in the second half of the
burst (Table 8).

However, it is evident from Figs. 14 and Tables 7 and 8 that as
the number of pulses in a burst increases, the median values of
the rise and fall times, FWHM and time intervals all decrease and
by about the same amount.  The GRBs with more pulses also have on
average significantly longer durations, higher fluences and
hardness ratios (Table~7). The variability index of a GRB was
taken to be the number of pulses $\geq 5 \sigma$ divided by the
time the GRB emission was also $\geq 5 \sigma$. The median values
are given in Table~9 for the four GRB categories. The GRBs with
more pulses have a higher variability index.  These results
provide an interesting interpretation of the two correlations
that have been reported for GRBs with known redshift: (1)\, the
more luminous GRBs to be more variable \citep{feniram:2001} and
(2)\, there is an anticorrelation between the arrival times of
high energy and low energy pulses in GRBs
\citep{nmb:2000,sal:2000}. Recently \citet{schaefer:2001} showed
that there is a relationship between the variability and spectral
lag \citep{ioka:2001}.  There is a good correlation between the
values of the variability obtained here and those of
\citet{feniram:2001}. GRBs with higher values of HR have lower
values of $<$V/V$_{\rm max}>$ implying they are a more distant
and luminous population \citep{schmidt:2001}.

Our knowledge on the shape of the emitting region in GRBs is
restricted because, due to relativistic beaming, only a small
portion of angular size \(\sim \Gamma^{-1}\) is visible to the
observer. Thus the observer is unable to distinguish a sphere
from a jet as long as \(\Gamma > \theta^{-1}\) where $\theta$ is
the radius of the opening angle of jet \citep{rhoads:1997}.
However as the source continues its rapid expansion, $\Gamma$
will decrease, and when \(\Gamma < \theta^{-1}\) there will be a
marked decrease in the observed flux. The steep time dependence
of the afterglow emission, sometimes with changes in the slope of
the spectrum, and radio emission have been widely interpreted as
evidence for emission from jets
\citep{castro:2001,mes:2001,frail:2001}.} The GRBs with more
pulses appear to have higher values of the Lorentz factor
$\Gamma$.  The higher values of $\Gamma$ may come from a more
efficient and more active central engine.  In one variation of
the internal shock model, it was assumed that the degree of
collimation of the jet depended on the mass M at the explosion
{\citep{kobay:2001}.  A wide jet involves a large mass that
results in a flow with a lower $\Gamma$.  The pulse properties
depend strongly on $\Gamma_{\rm min}, \Gamma_{\rm max}$ and the
radius of the photosphere R$\pm$.  GRBs with faster pulses
originate in collisions above R$\pm$ whereas GRBs with slower
pulses have smaller values of $\Gamma$ and some collisions below
the photosphere.  While this homogeneous model may explain pulse
properties in GRBs, the strong possibility of inhomogeneous jets
with a variable $\Gamma$ should also be examined.

Baryon loading can be a major problem in GRB models and severely
limit the attainable value of $\Gamma$
\citep{rees:1999,mrw:1998,sal:2000}. There maybe a broad range of
$\Gamma$'s in the outflow with the highest value occurring close
to the rotation axis where the baryon contamination should be at
a minimum.  At larger angles from the axis, there may be an
increasing degree of contamination with a corresponding drop in
$\Gamma$. The outcome of a collapse in a massive star whose iron
core collapsed to a black hole have been computed
\citep{macfad:1999}. The resulting jet that drives out through the
star is probably powered by a MHD process which can in principle
convert a large portion of the binding energy at the last stable
orbit into jet energy. The large amount of energy dumped into the
natural funnel-shaped channel creates a highly collimated jet,
focused into a small region of the sky. The largest value of
$\Gamma$ occurs on axis and decreases with increasing $\theta$
because the material coming at the observer has less energy at
larger angles. The emission is still beamed into an angle
$\Gamma^{-1}$ but in this inhomogeneous model the angle varies
across the opening angle of the jet \citep{rossi:2001}.  In this
situation the properties of the pulses in GRBs can be influenced
by the jet. The BATSE sample of the brightest GRBs should contain
a range of angles within the jet and hence different values of
$\Gamma$.  In this context it is reasonable to identify the
complex GRBs with more pulses and higher values of $\Gamma$ with
angles near the axis of jet. The GRBs that are viewed at larger
angles from the jet axis have on average lower values of
$\Gamma$, and develop at the greater distances from the central
engine and should have slower pulses.

In this context it is interesting to note that the pulse
evolution consisting of slower rise times and faster fall times in
the first half, is more pronounced for GRBs in category N than
either O or P. In a jet model with a variable $\Gamma$, the GRBs
in category N would be on average farther from the axis and more
sensitive to a clearing out effect such as additional baryon
loading or Compton drag \citep{rees:1999,ghis:2001} in the
initial phase of the GRB.

The steep and variable slope of the decay of GRB afterglows have
been widely interpreted as evidence for jets in GRBs
\citep{mes:2001,castir:2001}. If the axis of the jet is pointed
close to the observer, the GRB will be intense and the afterglow
should contain evidence for good alignment.  It is interesting
that the two brightest GRBs detected by WFC on BeppoSAX also were
the best aligned.  The recent detection of a bright GRB with a
fluence of \(10^{-4}\) ergs cm$^{-2}$ also had a very bright
afterglow \citep{castir:2001}. However many more GRBs and
afterglows are required to verify the existence of a pattern
between the strongest GRBs and their afterglows
\citep{frail:2001}. The distribution of the number of pulses per
GRB (Fig.~5) may broadly represent the beaming by the jet because
bursts with large numbers of pulses and higher variability
(Table~7) may be close to the axis and bursts with smaller
numbers of pulses and less variability further off-axis.

\section{Conclusions}

The properties of the brightest 319 GRBs in the BATSE current
catalogue have been analysed.The automatic pulse selection
process detected more than  3300 pulses. The distributions of
pulse rise and fall times, FWHM, areas, amplitudes and time
intervals between pulses are reasonably consistent with the
lognormal distribution. GRB pulse profiles can be elegantly
described by a small number of parameters that may be very useful
for simulations.  The lognormal distribution depends on the
product of probabilities arising from a combination of
independent events and these conditions must therefore apply to
the generation of the temporal and spectral properties of GRB
pulses. A wide range of burst parameters and also pulse
parameters were correlated and the results follow the trend
expected from the internal shock model. The pulse amplitude is
strongly anticorrelated with the other pulse timing parameters.
The time intervals between pulses and pulse amplitudes are
correlated with each other.

A comprehensive analysis has been performed between the first half
and second half of GRBs in three categories defined in terms of
N. No major differences were found between the distribution of
pulse properties between the first half and second half of the
GRBs.  There is a strong tendency for pulses to have slower rise
times and faster fall times in the first half of the burst.  This
trend is stronger in GRBs with small numbers of pulses.  The
pulse timing parameters and time intervals all decrease with
increase in N.  These results seem to be compatible with jet
models with either a $\Gamma$ that varies with the opening angle
or is constant and varies with the mass.   If $\Gamma$ varies with
the opening angle of the jet, the GRBs with higher values of
$\Gamma$ and greater variability are observed close to the axis of
the jet while GRBs with smaller number of pulses and less
variability are observed at larger angles from the jet.  Jets
with values of $\Gamma$ that vary with angle or with mass may
explain the luminosity-variability correlation and the
luminosity-energy lag correlation in GRBs with known redshift.

This study of the number of pulses in GRBs and their time
structure provides strong evidence for rotation powered systems
with intense magnetic fields and the added complexity of a jet.
These results can be well interpreted by internal shocks in the
framework of theoretical models for the formation of black holes
and subsequent jet formation.
\bibliography{katmonic,centeng_ch}
\bibliographystyle{apj}
\end{document}